\documentclass[journal]{IEEEtran}
\usepackage{verbatim}  
\usepackage{mathrsfs}  
\usepackage{bm}        
\usepackage{chngpage}  
\usepackage{array}
\usepackage{url}       
\usepackage{booktabs}  
\usepackage{multirow}
\usepackage{textcomp}  
\usepackage{setspace}
\usepackage{cite}
\usepackage{amsmath}
\usepackage{cleveref}  
\usepackage{amssymb}   
\usepackage{makecell}
\usepackage{graphicx}
\usepackage{xcolor}
\usepackage{epstopdf}
\usepackage{etoolbox}
\usepackage{ulem}

\makegapedcells
\setcellgapes{0pt}

\ifCLASSOPTIONcompsoc
	\usepackage[caption=false,font=normalsize,labelfont=sf,textfont=sf,subrefformat=parens,labelformat=parens]{subfig}
\else
	\usepackage[caption=false,font=footnotesize,subrefformat=parens,labelformat=parens]{subfig}
\fi

\makeatletter
\patchcmd{\@maketitle}
{\addvspace{0.5\baselineskip}\egroup}
{\addvspace{-1.0\baselineskip}\egroup}
{}
{}
\makeatother

\begin{document}
\title{Efficient Robust Dispatch of\\Combined Heat and Power Systems}

\author{
Yibao~Jiang,~\IEEEmembership{Student Member,~IEEE,}
Can~Wan,~\IEEEmembership{Member,~IEEE,}
Audun~Botterud,~\IEEEmembership{Member,~IEEE,}
Yonghua~Song,~\IEEEmembership{Fellow,~IEEE,} 
and Zhao~Yang~Dong,~\IEEEmembership{Fellow,~IEEE} 
\thanks{Y. Jiang and C. Wan are with the College of Electrical Engineering, Zhejiang University, Hangzhou 310027, China (e-mail: jiangyb@zju.edu.cn, canwan@zju.edu.cn. Corresponding author: Can Wan.)}
\thanks{A. Botterud is with the Laboratory for Information and Decision Systems, Massachusetts Institute of Technology, Cambridge, MA 02139 USA (e-mail: audunb@mit.edu).}
\thanks{Y. Song is with the Department of Electrical and Computer Engineering, University of Macau, Macau, China and also with the College of Electrical Engineering, Zhejiang University, Hangzhou 310027, China (e-mail: yhsongcn@zju.edu.cn).}
\thanks{Z. Y. Dong is with the School of Electrical Engineering and Telecommunications, University of New South Wales, Sydney, NSW 2052, Australia (e-mail: zydong@ieee.org).}
}


\maketitle

\normalem

\begin{abstract}
Combined heat and power systems facilitate efficient interactions between individual energy sectors for higher renewable energy accommodation. However, the feasibility of operational strategies is difficult to guarantee due to the presence of substantial uncertainties pertinent to renewable energy and multi-energy loads. This paper proposes a novel efficient robust dispatch model of combined heat and power systems based on extensions of disturbance invariant sets. The approach has high computational efficiency and provides flexible and robust strategies with an adjustable level of conservativeness. In particular, the proposed robust dispatch method obtains operational strategies by solving a nominal uncertainty-free dispatch problem, whose complexity is identical to a deterministic problem. The robustness against uncertainties is enhanced by endowing the nominal dispatch model with properly tightened constraints considering time-variant uncertainty sets. Towards this end, a novel direct constraint tightening algorithm is developed based on the dual norm to calculate multi-period tightened constraints efficiently without linear programming iterations. Furthermore, the budget uncertainty set is newly combined with constraint tightening to flexibly adjust the conservativeness level of robust solutions. The effectiveness of the proposed robust method is demonstrated in simulation studies of a test system in terms of computational efficiency, decision robustness and cost optimality.
\end{abstract}



\vspace{-2mm}
\section{Introduction}
\IEEEPARstart{I}{ncreasing} deployment of combined heat and power plants (CHPs) and distributed energy resources intensifies interactions between electric power systems and district heating systems. Combined heat and power systems (CHPS) are promising alternatives for enhancing reliable energy supply and promoting renewable integration \cite{mancarella2014mes,dall2017unlocking} via efficient coordination of different energy sectors \cite{jiang2019hybrid}.

Previous studies of CHPS focus on the economic dispatch of co-generation systems to optimally fulfill electricity and heat loads at the plant level \cite{guo1996algorithm,vasebi2007combined}. The combined heat and power dispatch problem is further studied by incorporating network-constrained energy flow \cite{liu2016combined} and temperature dynamics of heating systems \cite{li2016combined}. Electric boilers and thermal tanks are also exploited for efficient integration of wind power \cite{chen2015increasing}. There are also studies investigating benefits of combined heat and power dispatch by utilizing thermal inertia of heating systems \cite{huang2018heat,zhang2019day,li2015transmission}. In general, these studies focus on deterministic operational strategies for CHPS and perform well when renewable generation and loads follow forecasted values exactly. However, growing penetration of renewable energy sources introduces significant uncertainties to system modeling and operation, which may render deterministic solutions infeasible. The interdependence between electricity and heat systems also intensifies the complexity since both electricity and heat loads induce extra disturbances to CHPS. 

Therefore, a fundamental challenge in managing CHPS is to develop a robust dispatch method for operational decisions under uncertainty. 
A robust operational model based on linear decision rules is proposed for CHPs with maximization of electricity revenue in the day-ahead market \cite{zugno2014robust}. Robust management models for CHP-based microgrids and energy hubs are studied in \cite{nazari2018robust} and \cite{parisio2012robust}, respectively. An information gap decision method is utilized in \cite{aghaei2017optimal} to derive risk-averse and risk-taking operational strategies for combined heat and power plants. There are also studies focusing on robust scheduling of integrated electricity, heat and gas systems \cite{martinez2013robust,bai2017robust,cesena2018energy}. In \cite{martinez2013robust}, the robust counterpart of the original scheduling model is generated by introducing scenario sets. Security constraints of electricity transmission lines and gas pipelines are incorporated in \cite{bai2017robust} using a min-max paradigm. Reference \cite{cesena2018energy} proposes a two-stage iterative algorithm for robust optimization of smart multi-energy districts.
Some aforementioned robust dispatch methods are developed based on min-max schemes \cite{zugno2014robust,nazari2018robust,bai2017robust}. However, the number of variables and constraints may grow tremendously by introducing robust counterparts or scenarios, which raises severe challenges to the computational efficiency.
Meanwhile, operational strategies generated by the traditional min-max approach may be conservative due to the optimization under worst-case scenarios. Finally, previous works generally model the CHPS without including network-constrained electricity and heating flows \cite{zugno2014robust,nazari2018robust,aghaei2017optimal}.

In this paper, an efficient robust dispatch (ERD) model of CHPS is developed based on extensions of disturbance invariant sets to provide feasible operational strategies with enhanced computational efficiency and adjustable conservativeness levels. The proposed ERD method ensures the feasibility of operational strategies despite uncertainties pertaining to volatile renewable energy and loads by endowing a \emph{nominal} (i.e., uncertainty-free) dispatch problem with specific tightened constraints. Distinct from previous min-max algorithms that utilize robust counterparts for model reformulation, the proposed ERD method enhances robustness by solving a deterministic problem without introducing extra variables, ensuring high computational efficiency. Besides, the ERD model also incorporates a state-feedback policy to actively counteract disturbance effects with recourse actions.

The tightened constraints of the proposed ERD model are calculated based on disturbance invariant sets, which characterize the deviation of system states induced by disturbances and play a fundamental role in performance analysis and control synthesis \cite{Langson2004Robust}. The analysis of disturbance invariant sets normally models disturbances as time-invariant compact sets \cite{Mayne2001Robustifying,Langson2004Robust}, i.e., the upper and lower bounds of uncertainty parameters are constant over time. However, this is not applicable since uncertainty sets of renewable generation and loads obviously vary with time. For instance, the uncertainty level of PV outputs at noon is much higher than other periods. Besides, disturbance invariant sets are generally calculated based on iterations of linear programming \cite{kolmanovsky1998theory,rakovic2005invariant}, leading to heavy computation. Moreover, modeling uncertainties with the infinity norm (or the so-called ``box'' uncertainty set) is conservative since uncertainties at each time instant are not likely to reach their extreme values simultaneously. Therefore, the concept of disturbance invariant sets is extended based on set-theoretic analysis to make it compatible with time-variant ``heterogeneous'' uncertainty sets. More specifically, a direct constraint tightening algorithm based on the dual norm is developed to compute \emph{multi-period tightened constraints} efficiently without linear programming iterations. Last but not least, the ERD method incorporates the budget uncertainty set with direct constraint tightening to reduce conservativeness.

The main contributions are summarized as follows: 1) A novel efficient robust dispatch model of CHPS is proposed based on extension of disturbance invariant sets considering renewable power and load uncertainties, of which the robustness and computational efficiency are enhanced by solving a nominal uncertainty-free problem with multi-period tightened constraints.
2) The budget uncertainty set is newly combined with constraint tightening to flexibly adjust the level of conservativeness of the robust solutions.
3) A direct constraint tightening algorithm based on the dual norm is developed to efficiently derive multi-period tightened constraints considering time-variant uncertainty sets.
4) The proposed ERD method comprehensively models CHPS, including voltage deviations in electric power systems and temperature dynamics in district heating systems.

\vspace{-4mm}
\section{Modeling of CHPS}
\label{sec:model}
\vspace{-2mm}
\subsection{Distributed Energy Resources}
\vspace{-2mm}
Back-pressure CHPs are critical components with capability of supplying both electrical and thermal loads, expressed as
\begin{equation}
\eta^\text{CHP}_c P^\text{CHP}_c(t) = H^\text{CHP}_c(t), c\in\mathcal{C}, 
\end{equation}
\begin{equation}
\label{equ:chpminmax}
\begin{aligned}
& P_{c,\text{min}}^\text{CHP} \leq P^\text{CHP}_c(t) \leq P_{c,\text{max}}^\text{CHP}, |\Delta P^\text{CHP}_c(t)| \leq \Delta P_{c,\text{max}}^\text{CHP}, \\
& Q_{c,\text{min}}^\text{CHP} \leq Q^\text{CHP}_c(t) \leq Q_{c,\text{max}}^\text{CHP}, |\Delta Q^\text{CHP}_c(t)| \leq \Delta Q_{c,\text{max}}^\text{CHP},
\end{aligned}
\end{equation}
where $P^\text{CHP}_c$ and $Q^\text{CHP}_c$ denote the active and reactive power outputs of the $c$-th CHP constrained within $[P_{c,\text{min}}^\text{CHP},P_{c,\text{max}}^\text{CHP}]$ and $[Q_{c,\text{min}}^\text{CHP},Q_{c,\text{max}}^\text{CHP}]$, respectively; $H^\text{CHP}_c$ indicates the heating outflow; $\eta^\text{CHP}_c$ represents the power-to-heat ratio; $\Delta P^\text{CHP}_c$ and $\Delta Q^\text{CHP}_c$ are active and reactive ramping power with limits of $\Delta P_{c,\text{max}}^\text{CHP}$ and $\Delta Q_{c,\text{max}}^\text{CHP}$, respectively; $\mathcal{C}$ is the set of CHPs.

Heat pumps (HPs) consume electricity to supply heat with advantages of high efficiency and reliability, given as
\begin{gather}
\eta^\text{HP}_h P^\text{HP}_h(t) \!=\! H^\text{HP}_h(t), Q^\text{HP}_h(t) \!=\!\! \sqrt{1\!-\!(\delta^\text{HP}_h)^2} P^\text{HP}_h(t) / \delta^\text{HP}_h, \\
\label{equ:hpminmax}
P_{h,\text{min}}^\text{HP} \leq P^\text{HP}_h(t) \!\leq\! P_{h,\text{max}}^\text{HP}, |\Delta P^\text{HP}_h(t)| \!\leq\! \Delta P_{h,\text{max}}^\text{HP},h\!\in\!\mathcal{H},
\end{gather}
where $P^\text{HP}_h$ and $Q^\text{HP}_h$ are the active and reactive power inputs of the $h$-th HP; $H^\text{HP}_h$ is the thermal output; $P_{h,\text{max}}^\text{HP}$ and $P_{h,\text{min}}^\text{HP}$ denote maximum and minimum active power inputs, respectively; $\eta^\text{HP}_h$ is the power-to-heat ratio; $\delta^\text{HP}_h$ represents the power factor; the active ramping power $\Delta P^\text{HP}_h$ has an upper limit of $\Delta P_{h,\text{max}}^\text{HP}$; $\mathcal{H}$ is the set of HPs.

Battery units (BUs) can absorb or release electric power to maintain power balance, modeled as
\begin{gather}
\label{equ:buenergy}
E^\text{BU}_b(t+1) = \zeta^\text{BU}_b E^\text{BU}_b(t) + \big[\Delta T \eta^\text{BU}_b P^\text{BU}_b(t)\big]/C^\text{BU}_b, \\
\label{equ:buenergyminmax}
E_{b,\text{min}}^\text{BU}\leq E^\text{BU}_b(t) \leq E_{b,\text{max}}^\text{BU}, \\
\label{equ:buminmax}
P_{b,\text{min}}^\text{BU} \leq P^\text{BU}_b(t) \leq P_{b,\text{max}}^\text{BU},
|\Delta P^\text{BU}_b| \leq \Delta P_{b,\text{max}}^\text{BU}(t), b\in\mathcal{B},
\end{gather}
where $E^\text{BU}_b$ is the state of charge; $\zeta^\text{BU}_b$ represents the self-discharging coefficient; $P^\text{BU}_b$ denotes charging or discharging power; $\Delta T$ indicates the time step; $\eta^\text{BU}_b$ is the charging or discharging efficiency, expressed as $\eta^\text{BU}_b\!\!\!=\!\eta_b^\text{ch},P^\text{BU}_b\!\geq\! 0$, and $\eta^\text{BU}_b\!\!\!=\! 1/\eta_b^\text{dch},P^\text{BU}_b\!<\! 0$, respectively; $C^\text{BU}_b$ is the rated capacity; $E^\text{BU}_b$ is maintained within $[E_{b,\text{min}}^\text{BU},E_{b,\text{max}}^\text{BU}]$ to avoid over-discharge/charge; $P_{b,\text{min}}^\text{BU}/P_{b,\text{max}}^\text{BU}$ are rated discharging/charging power; $\Delta P_{b,\text{max}}^\text{BU}$ is the ramping limit; $\mathcal{B}$ is the set of BUs.

The model of thermal storage tanks (TSs) is given as
\begin{gather}
\label{equ:tsenergy}
E^\text{TS}_s(t+1) = \zeta^\text{TS}_s E^\text{TS}_s(t) + \big[\Delta T\eta^\text{TS}_s H^\text{TS}_s(t)\big]/C^\text{TS}_s, \\
\label{equ:tsenergyminmax}
E_{s,\text{min}}^\text{TS}\leq E^\text{TS}_s(t) \leq E_{s,\text{max}}^\text{TS}, \\
\label{equ:tsminmax}
H_{s,\text{min}}^\text{TS} \!\leq\! H^\text{TS}_s(t) \!\leq\! H_{s,\text{max}}^\text{TS},
|\Delta H^\text{TS}_s(t)| \!\leq\! \Delta H_{s,\text{max}}^\text{TS}, s\!\in\!\mathcal{S},
\end{gather}
where $E^\text{TS}_s$ is the thermal storage level; $\zeta^\text{TS}_s$ represents the coefficient accounting for internal heat loss; $H^\text{TS}_s$ is the heating inflow or outflow; the heating efficiency is formulated as $\eta^\text{TS}_s\!\!\!=\!\eta_s^\text{ch},H^\text{TS}_s\!\geq\! 0$, and $\eta^\text{TS}_s\!\!\!=\! 1/\eta_s^\text{dch},H^\text{TS}_s\!<\! 0$, respectively; $C^\text{TS}_s$ indicates the thermal storage capacity; the thermal storage level is restricted by $[E_{s,\text{min}}^\text{TS},E_{s,\text{max}}^\text{TS}]$; $H_{s,\text{min}}^\text{TS}$ and $H_{s,\text{max}}^\text{TS}$ denote discharging and charging limits, respectively; $\Delta H_{s,\text{max}}^\text{TS}$ is the heat ramping limit; $\mathcal{S}$ is the set of TSs.

\vspace{-5mm}
\subsection{Power Distribution Network}
\vspace{-2mm}
The electric power flow is derived based on nodal power balance and branch equations. Let $\dot{\boldsymbol{S}}$ and $\dot{\boldsymbol{V}}$ denote complex nodal power injections and nodal voltages, respectively. Then the complex branch current $\dot{\boldsymbol{I}}_B$ can be formulated as
\begin{equation}
\boldsymbol{A} \dot{\boldsymbol{I}}_B = \dot{\boldsymbol{I}} = \tilde{\boldsymbol{S}}/\tilde{\boldsymbol{V}},
\end{equation}
where $\boldsymbol{A}$ represents the incidence matrix of the power network; $A_{ij}\!=\!1$ if bus $i$ is the ``from'' bus of branch $j$, and -1 otherwise; $\dot{\boldsymbol{I}}$ is the nodal current injection; $\tilde{ }$ denotes the complex conjugate. Active power flow $\boldsymbol{T}$ can be approximated by linear equations \cite{jiang2019stochastic,geidl2007optimal}, shown as 
\begin{gather}
\label{equ:transcon}
\boldsymbol{T}\mathrel{\mathop:}=[T_l,l\in\mathcal{L}^\text{E}]=\Re\big[\hat{\boldsymbol{A}}^\intercal\dot{\boldsymbol{V}}\odot\tilde{\boldsymbol{I}}_B\big], \\
\label{equ:transconminmax}
-\boldsymbol{T}_\text{max} \leq \boldsymbol{T}  \leq \boldsymbol{T}_\text{max},
\end{gather}

\noindent where $T_l$ represents active power flow of branch $l$; $\mathcal{L}^\text{E}$ denotes the set of lines in the electric power network; power flow limits are denoted as $\boldsymbol{T}_\text{max}$; matrix $\hat{\boldsymbol{A}}$ is derived from $\boldsymbol{A}$, where $\hat{A}_{ij}\!\!=\!\!1$ if $A_{ij}\!\!=\!\!1$, and 0 otherwise; $\Re[\cdot]$ refers to the real part of a complex number; operator $\odot$ represents element-wise product of two vectors. A linear voltage dynamic model based on Z-bus sensitivities is utilized to characterize voltage deviations with respect to variations of power injections \cite{jiang2019stochastic, valverde2013model}. The voltage sensitivity of bus $i$ with respect to the active power injection at bus $k$ is obtained by
\begin{gather}
\label{equ:volp}
\frac{\partial V_i}{\partial P_k} = \frac{1}{V_i} \Re\left(\tilde{V}_i \frac{\partial \dot{V}_i}{\partial P_k} \right), \forall i, k \in \mathcal{N}^\text{E}, \\
\label{equ:volpcomp}
\frac{\partial \dot{V}_i}{\partial P_k} = \sum_{j\in\mathcal{N}^\text{E}} \frac{-\dot{Z}_{ij}\tilde{S}_j}{\tilde{V}_j^2} \frac{\partial \tilde{V}_j}{\partial P_k} + \frac{\dot{Z}_{ik} \tilde{V}_k}{\tilde{V}_k^2}, 
\end{gather}
where $V_i$ and $P_k$ denote the voltage magnitude of bus $i$ and the active power injection at bus $k$, respectively; $\dot{Z}_{ij}$ is the complex branch impedance between bus $i$ and $j$; $\mathcal{N}^\text{E}$ is the set of buses in the electric power network. The derivative of voltage magnitude with respect to the reactive power injection can be derived similarly and is not shown here for simplicity.
Let $V_i^\text{min}$ and $V_i^\text{max}$ denote the minimum and maximum voltage magnitudes at bus $i$, the voltage limit is shown as
\begin{equation}
\label{equ:vol}
V_i^\text{min} \leq V_i \leq V_i^\text{max}, \forall i\in\mathcal{N}^\text{E}.
\end{equation}

\vspace{-6mm}
\subsection{District Heating Network}
\vspace{-2mm}
Temperature changes are transferred slowly in the heating network, and it is important to model heat transport delays from sources to sinks. The heating system is assumed to operate at constant flow to ensure stable hydraulic conditions \cite{li2015transmission}. Then the time delay $\tau_l(t)$ describing time steps of heat delivery in pipe $l$ at time $t$ can be determined by mass flow rates based on the node method \cite{palsson1999equivalent}, expressed as
\begin{equation}
\tau_l(t) \!=\! \min\Big\{\bar{\tau}\!\geq\! 0:\!\!\!\sum_{\tau=t-\bar{\tau}}^t\!\!\! m_l(\tau)\Delta T \!>\! \frac{\pi L_l D^2_l \rho}{4} \Big\},l\!\in\!\mathcal{L}^\text{H},
\end{equation}
where $m_l(\tau)$ is the predefined mass flow rate of pipe $l$ at time $\tau$; $L_l$ and $D_l$ denote the length and the diameter of pipe $l$; $\mathcal{L}^\text{H}$ represents the set of pipelines; $\rho$ is the water density.
The heat input at the source node $i$ in the heating network is given by
\begin{equation}
\label{equ:heatinput}
\begin{aligned}
\sum_{c\in\mathcal{C}_i} H^\text{CHP}_c(t)&+ \sum_{h\in\mathcal{H}_i} H^\text{HP}_h(t) - \sum_{s\in\mathcal{S}_i} H^\text{TS}_s(t) \\
&= c^\text{w} m^\text{in}_i(t) \left(T^\text{s}_i(t)-T^\text{r}_i(t)\right),
\end{aligned}
\end{equation}
where $\mathcal{C}_i$, $\mathcal{H}_i$ and $\mathcal{S}_i$ denote sets of CHPs, HPs and TSs at node $i$ in the district heating network, respectively; $c^\text{w}$ is the specific heat capacity of water; $m^\text{in}_i$ is the mass flow injection at node $i$; $T^\text{s}_i$ and $T^\text{r}_i$ indicate supply temperature and return temperature at node $i$, respectively. 
Similarly, the heat output at the load node $i$ is formulated as
\begin{equation}
\label{equ:heatload}
H^\text{D}_i(t) = c^\text{w}m^\text{ot}_i(t)\left(T^\text{s}_i(t)-T^\text{r}_i(t)\right),
\end{equation}
where $H^\text{D}_i$ is the heat demand; $m^\text{ot}_i$ is mass outflow at node $i$. 

The heat delivery is characterized by the mapping relationship between inlet and outlet pipe temperatures, given by
\begin{equation}
\label{equ:suptemp}
T^\text{s}_j(t) = T^\text{g}(t)+\left(T^\text{s}_i(t-\tau_l(t))-T^\text{g}(t)\right)\times \text{e}^{-\frac{k_l\Delta T\tau_l(t)}{A_l\rho c^\text{w}}},
\end{equation}
\begin{equation}
\label{equ:rettemp}
T^\text{r}_i(t) = T^\text{g}(t)+\left(T^\text{r}_j(t-\tau_l(t))-T^\text{g}(t)\right)\times \text{e}^{-\frac{k_l\Delta T\tau_l(t)}{A_l\rho c^\text{w}}},
\end{equation}
where $T^\text{g}$ is the ground temperature; $k_l$ and $A_l$ denote heat conductivity and cross-section area of pipe $l$; $i$ and $j$ are ``from'' and ``to'' nodes of pipe $l$ with respect to the supply heating network. The supply and return temperatures are confined by desired bounds, expressed as 
\begin{equation}
\label{equ:tempcon}
T^\text{s}_{i,\text{min}}\leq T^\text{s}_i(t)\leq T^\text{s}_{i,\text{max}},T^\text{r}_{i,\text{min}}\leq T^\text{r}_i(t)\leq T^\text{r}_{i,\text{max}}.
\end{equation}

Electric power and district heating balances are given as
\begin{gather}
\label{equ:activebal}
\sum_{c\in\mathcal{C}}\!\! P^\text{CHP}_c \!+\! P^\text{G}_g \!+\! \sum_{p\in\mathcal{P}}\!\! P^\text{PV}_p \!\!=\!\! \sum_{i\in\mathcal{N}^\text{E}}\!\! P^\text{D}_i \!+\! \sum_{b\in\mathcal{B}}\!\! P^\text{BU}_b \!+\! \sum_{h\in\mathcal{H}}\!\! P^\text{HP}_h, \\
\label{equ:reactivebal}
\sum_{c\in\mathcal{C}}Q^\text{CHP}_c + Q^\text{G}_g = \sum_{i\in\mathcal{N}^\text{E}}Q^\text{D}_i + \sum_{h\in\mathcal{H}}Q^\text{HP}_h, \\
\label{equ:heatbal}
\sum_{c\in\mathcal{C}}H^\text{CHP}_c + \sum_{h\in\mathcal{H}}H^\text{HP}_h = \sum_{i\in\mathcal{N}^\text{H}}H^\text{D}_i + \sum_{s\in\mathcal{S}}H^\text{TS}_s,
\end{gather}
where $P^\text{G}_g$ and $Q^\text{G}_g$ are the exchange active and reactive power with the main power grid; $P^\text{D}_i$ and $Q^\text{D}_i$ indicate active and reactive electricity loads, respectively; $P^\text{PV}_p$ refers to the power output of photovoltaic panels (PVs); $\mathcal{P}$ is the set of PVs; $\mathcal{N}^\text{H}$ indicates the set of nodes in the heating network. Power exchange $P^\text{G}_g$ and $Q^\text{G}_g$ are restricted in intervals to avoid power fluctuation affecting the main grid, indicated by
\begin{equation}
\label{equ:gridcon}
P_{g,\text{min}}^\text{G}\leq P^\text{G}_g(t)\leq P_{g,\text{max}}^\text{G}, Q_{g,\text{min}}^\text{G}\leq Q^\text{G}_g(t)\leq Q_{g,\text{max}}^\text{G}.
\end{equation}

Electricity loads, heat loads and PV power are represented by prediction intervals \cite{wan2017probabilistic}, expressed as
\begin{gather}
\label{equ:elecdis}
P_{i,\text{min}}^\text{D}\!(t) \!\!\leq\!\! P^\text{D}_i\!(t) \!\leq\! P_{i,\text{max}}^\text{D}\!(t),\! Q_{i,\text{min}}^\text{D}\!(t) \!\leq\! Q^\text{D}_i\!(t) \!\leq\! Q_{i,\text{max}}^\text{D}\!(t), \\
\label{equ:heatpvdis}
H_{i,\text{min}}^\text{D}\!(t) \!\!\leq\!\! H^\text{D}_i\!(t) \!\!\leq\!\! H_{i,\text{max}}^\text{D}\!(t),\!P_{p,\text{min}}^\text{PV}\!(t) \!\!\leq\!\! P^\text{PV}_p\!(t) \!\!\leq\!\! P_{p,\text{max}}^\text{PV}\!(t).
\end{gather}

\section{Compact Formulation of ERD Model}

\subsection{Objective}
\vspace{-1mm}
The objective is to minimize total costs, including fuel and maintenance costs, and power exchange with the main grid,
\begin{equation}
\begin{aligned}
J \!=\!\! \sum_{t=0}^{T-1}\!\bigg\{\! & \sum_{c\in\mathcal{C}} \alpha^\text{CHP}_c P^\text{CHP}_c(t) \!+\! \sum_{b\in\mathcal{B}} \alpha^\text{BU}_b|P^\text{BU}_b(t)|+ \\
& \sum_{s\in\mathcal{S}}\!\! \alpha^\text{TS}_s|H^\text{TS}_s(t)| \!+\! \sum_{h\in\mathcal{H}}\!\! \alpha^\text{HP}_h P^\text{HP}_h(t) \!+\! \alpha^\text{G}_g P^\text{G}_g(t)\!\bigg\},
\end{aligned}
\end{equation}
where $T$ is the horizon length; $\alpha^\text{CHP}_c$ is the cost coefficient of CHPs; $\alpha^\text{BU}_b$, $\alpha^\text{TS}_s$, and $\alpha^\text{HP}_h$ denote maintenance cost factors for BUs, TSs, and HPs, respectively. $\alpha^\text{G}_g$ is the electricity price. 

\subsection{Constraints}
\vspace{-1mm}
The formulation in Section \ref{sec:model} can be regarded as a primitive form of constraints. A set-based representation is given here to make the problem formulation as compact as possible. Let $x(t)$, $u(t)$, $y(t)$ and $w(t)$ denote state, control, analysis and disturbance variables, respectively, defined as
\begin{gather}
x(t)\mathrel{\mathop:}=\big[E^\text{BU}_{b\in\mathcal{B}}(t),E^\text{TS}_{s\in\mathcal{S}}(t)\big]^\intercal, \\
u(t) \mathrel{\mathop:}= \big[P^\text{CHP}_{c\in\mathcal{C}}(t),Q^\text{CHP}_{c\in\mathcal{C}}(t),P^\text{G}_g(t),Q^\text{G}_g(t),P^\text{HP}_{h\in\mathcal{H}}(t)\big]^\intercal, \\
\begin{aligned}
& y(t) \mathrel{\mathop:}= \big[P^\text{BU}_{b\in\mathcal{B}}(t),H^\text{TS}_{s\in\mathcal{S}}(t),T_{l\in\mathcal{L}^\text{E}}(t),\\
& \qquad\qquad\qquad\qquad V_{i\in\mathcal{N}^\text{E}}(t),T^\text{s}_{i\in\mathcal{N}^\text{H}}(t),T^\text{r}_{i\in\mathcal{N}^\text{H}}(t)\big]^\intercal, 
\end{aligned} \\
w(t) \mathrel{\mathop:}= \big[P^\text{PV}_{p\in\mathcal{P}}(t),P^\text{D}_{i\in\mathcal{N}^\text{E}}(t),Q^\text{D}_{i\in\mathcal{N}^\text{E}}(t),H^\text{D}_{i\in\mathcal{N}^\text{H}}(t)\big]^\intercal,
\end{gather}
where the dimensions of state, control, analysis and disturbance variables are $n_x$, $n_u$, $n_y$ and $n_w$ respectively.
Then a linear, discrete-time state space model can be formulated as
\begin{gather}
\label{equ:comsys}
x(t+1) = A x(t)+B u(t)+D w(t), \\
\label{equ:comoutput}
y(t) = C u(t)+E w(t),
\end{gather}
where the dynamic of energy levels in storage units is characterized in (\ref{equ:comsys}); analysis variables are mapped with control actions and disturbances by (\ref{equ:comoutput}), derived from (\ref{equ:transcon}), (\ref{equ:volp}), (\ref{equ:volpcomp}), (\ref{equ:heatinput})-(\ref{equ:rettemp}) and (\ref{equ:activebal})-(\ref{equ:heatbal}).
Other constraints can be compactly formulated as polyhedral sets, expressed by
\begin{gather}
\label{equ:oriconone}
\left(x(t),u(t),y(t)\right)\in\mathbb{X}\times\mathbb{U}\times\mathbb{Y}, \\
\label{equ:oricontwo}
\left(\Delta u(t),\Delta y(t)\right)\in\Delta\mathbb{U}\times\Delta\mathbb{Y},
\end{gather}
where the state constraint $\mathbb{X}$ is derived from boundary limits on energy storage levels (\ref{equ:buenergyminmax}) and (\ref{equ:tsenergyminmax}); the control constraint $\mathbb{U}$ is confirmed by the CHP power generation capability (\ref{equ:chpminmax}), the HP power input limit  (\ref{equ:hpminmax}) and the exchange power limit with the main grid (\ref{equ:gridcon}); the constraint on the analysis variable $\mathbb{Y}$ is obtained from charging/discharging power limits (\ref{equ:buminmax}), (\ref{equ:tsminmax}), the power flow limit (\ref{equ:transconminmax}), the voltage constraint (\ref{equ:vol}) and the temperature limit (\ref{equ:tempcon}); ramping constraints on $u(t)$ and $y(t)$ are determined by (\ref{equ:chpminmax}), (\ref{equ:hpminmax}), (\ref{equ:buminmax}) and (\ref{equ:tsminmax}). 
For instance, constraints on state and control variables are represented by,
\begin{gather}
\mathbb{X}\!\mathrel{\mathop:}=\!\big\{x(t)\!:\![E^\text{BU}_{b,\text{min}},E^\text{TS}_{s,\text{min}}]^\intercal \!\leq\! x(t) \!\leq\! [E^\text{BU}_{b,\text{max}},E^\text{TS}_{s,\text{max}}]^\intercal \big\}, \\
\begin{aligned}
\mathbb{U}\!\mathrel{\mathop:}=\!\big\{\! u(t)\!:\![&P^\text{CHP}_{c,\text{min}},\!Q^\text{CHP}_{c,\text{min}},\!P^\text{G}_{g,\text{min}},\!Q^\text{G}_{g,\text{min}},\!P^\text{HP}_{h,\text{min}}]^\intercal \!\!\leq\! u(t) \\
&\!\leq\! [P^\text{CHP}_{c,\text{max}},\!Q^\text{CHP}_{c,\text{max}},\!P^\text{G}_{g,\text{max}},\!Q^\text{G}_{g,\text{max}},\!P^\text{HP}_{h,\text{max}}]^\intercal \!\big\}.
\end{aligned}
\end{gather}

The uncertainty $w(t)$ indicates actual PV power, electricity and heat demands, modeled by prediction intervals as
\begin{gather}
w(t)\in\mathbb{W}(t)\mathrel{\mathop:}=\big\{w(t):w_\text{min}(t)\leq w(t)\leq w_\text{max}(t)\big\}, \\
\left\{
\begin{aligned}
& w_\text{min} = [P^\text{PV}_{p,\text{min}}(t),P^\text{D}_{i,\text{min}}(t),Q^\text{D}_{i,\text{min}}(t),H^\text{D}_{i,\text{min}}(t)]^\intercal, \\
& w_\text{max} = [P^\text{PV}_{p,\text{max}}(t),P^\text{D}_{i,\text{max}}(t),Q^\text{D}_{i,\text{max}}(t),H^\text{D}_{i,\text{max}}(t)]^\intercal. 
\end{aligned}
\right.
\end{gather}

The min-max algorithms have been used to cope with the robust dispatch problem by introducing robust counterparts \cite{zugno2014robust,nazari2018robust,bai2017robust,parisio2012robust}. However, the problem size grows significantly with auxiliary variables and constraints, especially in a multi-period problem with network-constrained energy flow. To ensure the robustness with significant computational efficiency, a novel ERD model of CHPS is proposed based on extensions of disturbance invariant sets to obtain operational strategies by solving a nominal uncertainty-free dispatch problem without introducing any additional variables and constraints. 
In particular, a nominal disturbance-free system is defined as opposed to the realistic system (\ref{equ:comsys}) and (\ref{equ:comoutput}), given as
\begin{gather}
\label{equ:nomsys}
\bar{x}(t+1) = A \bar{x}(t)+B \bar{u}(t) + D \bar{w}(t), \\
\label{equ:nomoutput}
\bar{y}(t) = C \bar{u}(t) + E \bar{w}(t),
\end{gather}
where $\bar{x}(t)$, $\bar{u}(t)$ and $\bar{y}(t)$ are nominal uncertainty-free state, control and analysis variables, respectively; $\bar{w}(t)$ denotes expected prediction values of uncertainties, namely $\bar{w}(t)\!\mathrel{\mathop:}=\! [\bar{P}^\text{PV}_{p}(t),\bar{P}^\text{D}_{i}(t),\bar{Q}^\text{D}_{i}(t),\bar{H}^\text{D}_{i}(t)]^\intercal$. The uncertainty variable $w(t)$ is replaced by its predicted value in (\ref{equ:nomsys}) and (\ref{equ:nomoutput}) so that the ERD method solves a nominal problem. The computational complexity is significantly reduced compared to the traditional min-max algorithms, since no auxiliary variables and constraints are introduced. However, if nominal variables (i.e., $\bar{x}(t),\bar{u}(t),\bar{y}(t)$) are still restricted by original constraints (\ref{equ:oriconone}) and (\ref{equ:oricontwo}), the obtained solution is obviously not robust. As a consequence, more stringent constraints should be imposed on nominal variables to preserve the robustness, given as
\begin{gather}
\label{equ:tightconone}
\left(\bar{x}(t),\bar{u}(t),\bar{y}(t)\right)\in\bar{\mathbb{X}}(t)\times\bar{\mathbb{U}}(t)\times\bar{\mathbb{Y}}(t), \\
\label{equ:tightcontwo}
\left(\Delta\bar{u}(t),\Delta \bar{y}(t)\right)\in\Delta\bar{\mathbb{U}}(t)\times\Delta\bar{\mathbb{Y}}(t),
\end{gather}
where $\bar{\mathbb{X}}(t)$, $\bar{\mathbb{U}}(t)$, $\bar{\mathbb{Y}}(t)$, $\Delta\bar{\mathbb{U}}(t)$, and $\Delta\bar{\mathbb{Y}}(t)$ represent multi-period tightened constraints. The key issue is to properly derive these tightened constraints, as detailed in the next section.

\vspace{-4mm}
\section{Constraint Tightening of ERD Model}
\vspace{-1mm}
This section develops a rigorous method to determine multi-period tightened constraints based on set-theoretical analysis and extensions of disturbance invariant sets.
Some notations are given first for clarity. Set addition is defined by $\mathbb{A}\!+\!\mathbb{B}\!\mathrel{\mathop:}=\!\left\{a\!+\!b\!:\! a\!\in\!\mathbb{A},b\!\in\!\mathbb{B}\right\}$. The Minkowski set subtraction is defined by $\mathbb{A}\!\ominus\!\mathbb{B}\!\mathrel{\mathop:}=\!\left\{x\!:\! x\!+\!\mathbb{B}\!\subset\! \mathbb{A}\right\}$. $\|a\|_p$ denotes the $p$-norm of $a$. The mapping operator of sets is denoted as $M\mathbb{A}\mathrel{\mathop:}=\left\{Ma: a\in\mathbb{A}\right\}$.

\vspace{-4mm}
\subsection{Preliminaries of Disturbance Invariant Sets}
\vspace{-2mm}
\label{sec:dinvariant}
To actively counteract disturbance effects, an affine state-feedback control policy $\pi$ is utilized to map the control input with a nominal level and a state-feedback term, defined as
\begin{equation}
\label{equ:feedlaw}
\pi \mathrel{\mathop:} u(t)=\mu_t(x(t))=\bar{u}(t)\!+\!K(x(t)\!-\!\bar{x}(t)), \forall t\in\mathbb{I}_0^{T-1},
\end{equation}
where $\mathbb{I}_a^{b}$ represent the set of integers from $a$ to $b$; $K$ is the feedback gain. Substituting $u(t)$ in (\ref{equ:comsys}) with $\pi$ and subtracting (\ref{equ:nomsys}) from (\ref{equ:comsys}), an autonomous system can be obtained as
\begin{equation}
\label{equ:antsys}
\dot{x}(t+1) = \varPhi \dot{x}(t) + D \dot{w}(t),
\end{equation}
where $\dot{x}(t)\!=\!x(t)-\bar{x}(t)$, $\dot{w}(t)\!=\!w(t)-\bar{w}(t)$ and $\varPhi\!=\!A\!+\!BK$; $\dot{x}$ can be interpreted as the state deviation from the nominal level $\bar{x}$; $\dot{w}$ indicates the forecasting error and $\dot{w}(t)\!\in\!\dot{\mathbb{W}}(t)$. Note that the state space equation (\ref{equ:antsys}) captures the relationship between $\dot{w}(t)$ and $\dot{x}(t)$, then the definition of disturbance invariant sets \cite{Langson2004Robust} can be given if the disturbance $\dot{w}(t)$ is subject to a time-invariant uncertainty set $\dot{\mathbb{W}}$.

\emph{Definition 1}. $\mathcal{X}\subseteq\mathbb{R}^{n_x}$ is disturbance invariant for system (\ref{equ:antsys}), if $\varPhi \dot{x}+D \dot{w}\in\mathcal{X}$ for every $\dot{x}\in\mathcal{X}$ and every $\dot{w}\in\dot{\mathbb{W}}$. 

\vspace{-4mm}
\subsection{Definition of Multi-Period Tightened Constraints}
\vspace{-1mm}
Definition 1 is proposed based on time-invariant uncertainty sets, namely $\dot{w}(t)\in\dot{\mathbb{W}},\forall t\in\mathbb{I}_0^{T-1}$. However, it is not applicable here since prediction intervals of renewable power and loads are neither constant nor monotonic. If only the maximum uncertainty set is considered, the results of constraint tightening may be too conservative. Hence, the concept of disturbance invariant sets is extended by defining the \emph{multi-period 0-reachable set} $\mathcal{X}(t),t\in\mathbb{I}_1^T$, which is more accurate for describing time-variant uncertainty sets $\dot{\mathbb{W}}(t)$. 

\emph{Definition 2}. A set $\mathcal{X}(t)\subseteq\mathbb{R}^{n_x}$ is called the multi-period 0-reachable set at $t$ for the system (\ref{equ:antsys}), if $\mathcal{X}(t)$ satisfies the following two conditions: i) $\dot{x}(t)\in\mathcal{X}(t)$ for any disturbance sequence $\dot{\boldsymbol{w}} \mathrel{\mathop:}=\left\{\dot{w}(0),\dot{w}(1),\cdots,\dot{w}(T-1)\right\}\in\dot{\mathbb{W}}(0)\times\dot{\mathbb{W}}(1)\times\cdots\times\dot{\mathbb{W}}(T-1)$ with the initial state $\dot{x}(0)=\boldsymbol{0}$. ii) For any $\dot{x}(t)\in\mathcal{X}(t)$, there exist at least a disturbance sequence $\dot{\boldsymbol{w}}$ that ensures $\dot{x}(t)=\dot{\phi}(t;\boldsymbol{0},\dot{\boldsymbol{w}})$, where $\dot{\phi}(t;x_0,\dot{\boldsymbol{w}})$ represents the solution of the state space equation (\ref{equ:antsys}) at $t$ with initial state $x_0$ and a disturbance sequence $\dot{\boldsymbol{w}}$.

Based on Definition 2, $\mathcal{X}(t)$ can be expressed as 
\begin{equation}
\label{equ:multiset}
\mathcal{X}(t) \mathrel{\mathop:}=\sum_{i=0}^{t-1}\varPhi^{t-i-1}D\dot{\mathbb{W}}(i),t\in\mathbb{I}_1^T.
\end{equation}
The multi-period 0-reachable set $\mathcal{X}(t)$ provides a pointwise-in-time characterization of state deviations from $t\!=\!1$ to $t\!=\!T$, based on which the multi-period tightened constraints can be defined, given as the following proposition.

\emph{Proposition 1}. Suppose that the nominal control input $\bar{\boldsymbol{u}}$, analysis variable $\bar{\boldsymbol{y}}$, and state $\bar{\boldsymbol{x}}$ of uncertainty-free system (\ref{equ:nomsys}) satisfy following multi-period tightened constraints,
\begin{equation}
\label{equ:tightconsnew}
\left\{
\begin{aligned}
&\bar{u}(t)\in\bar{\mathbb{U}}(t)\mathrel{\mathop:}=\mathbb{U}\ominus K\mathcal{X}(t),\\
&\bar{y}(t)\in\bar{\mathbb{Y}}(t)\mathrel{\mathop:}=\mathbb{Y}\ominus CK\mathcal{X}(t)\ominus E\dot{\mathbb{W}}(t),\\
&\bar{x}(t)=\bar{\phi}(t;x_0,\bar{\boldsymbol{u}})\in\bar{\mathbb{X}}(t)\mathrel{\mathop:}=\mathbb{X}\ominus\mathcal{X}(t).
\end{aligned}
\right.
\end{equation}
Based on the feedback policy $\pi$ defined in (\ref{equ:feedlaw}), variables $\boldsymbol{x}$, $\boldsymbol{y}$ and $\boldsymbol{u}$ satisfy the following condition $\forall \dot{w}(t)\in\dot{\mathbb{W}}(t),t\!\in\!\mathbb{I}_0^{T-1}$, 
\begin{equation}
\label{equ:oriconsnew}
\left\{
\begin{aligned}
&u(t)=\bar{u}(t)+K(x(t)-\bar{x}(t))\in\mathbb{U},\\
&y(t)=\bar{y}(t)+CK(x(t)-\bar{x}(t))+E\dot{w}(t)\in\mathbb{Y},\\
&x(t)=\phi(t;x_0,\pi,\dot{\boldsymbol{w}})\in\mathbb{X}.
\end{aligned}
\right.
\end{equation}
Proposition 1 yields the fact that the satisfaction of tightened constraints (\ref{equ:tightconsnew}) is sufficient to guarantee feasibility. Thus, ERD enhances robustness by solving a disturbance-free problem with multi-period tightened constraints. 

\vspace{-4mm}
\subsection{Direct Constraint Tightening Algorithm}
\vspace{-2mm}
\label{sec:onestep}
One potential issue of introducing time-variant uncertainty sets is the increasing computation burden, since multi-period tightened constraints ($\bar{\mathbb{X}}(t),t\!\in\!\mathbb{T}_1^T$, respectively) of all time slots should be calculated. The computing time grows significantly if the iterative solution of linear programming \cite{kolmanovsky1998theory,rakovic2005invariant} is used.
In this paper, a new direct constraint tightening algorithm is developed based on the dual norm to calculate degrees of constraint restrictions without iterations. For conciseness, only the concrete algorithm for $\bar{\mathbb{X}}(t)$ is given here. $\mathbb{X}$ is given as
\begin{equation}
\label{equ:hrep}
\mathbb{X}\mathrel{\mathop:}=\big\{x\in\mathbb{R}^{n_x}: s^\intercal_i x\leq r_i,\forall i\in\mathbb{I}_1^M\big\},
\end{equation}
where $s_i$ and $r_i$ represent the left-hand coefficient and right-hand scalar of the $i$-th constraint, respectively; $M$ is the number of state constraints.
Recalling Proposition 1, the tightened state constraint $\bar{\mathbb{X}}(t)$ is derived by subtracting the multi-period 0-reachable set $\mathcal{X}(t)$ from $\mathbb{X}$, expressed as
\begin{equation}
\begin{aligned}
\bar{\mathbb{X}}(t)\mathrel{\mathop:}=\Big\{x\in\mathbb{R}^{n_x}:s^\intercal_i \big(x+\sum_{\tau=0}^{t-1}\varPhi^{t-\tau-1}D\dot{w}(\tau)\big)\leq r_i,&\\
\forall \dot{w}(\tau)\in\dot{\mathbb{W}}(\tau),\forall i\in\mathbb{I}_1^M\Big\}.&
\end{aligned}
\end{equation}
Multi-period tightened constraints are reformulated by introducing a normalized uncertainty vector $\tilde{w}\!\in\!\mathbb{R}^{T}$, shown as
\begin{equation}
\label{equ:statesup}
\begin{aligned}
\bar{\mathbb{X}}(t)\!\!\mathrel{\mathop:}=\!\!\Big\{\!&x\in\mathbb{R}^{n_x}: s^\intercal_i x\leq r_i-\\
&\!\sum_{j=1}^{n_w}\! \sup_{\!\tilde{w}}\{\varphi^{ij}_t W^j (\tilde{w}\!+\!\varpi_j)\!:\!\|\tilde{w}\|_\infty\!\!\leq\! 1\!\},\!\forall i\!\in\!\mathbb{I}_1^M\!\Big\},
\end{aligned}
\end{equation}
where $\sup\{\cdot\}$ is the abbreviation of ``supremum''; then $\sup_{\tilde{w}}\{\varphi^{ij}_t W^j \tilde{w}:\|\tilde{w}\|_\infty\leq 1\}$ defines a support function of $\varphi^{ij}_t W^j$ over the set $\{\|\tilde{w}\|:\|\tilde{w}\|_\infty\leq 1\}$; $\varphi^{ij}_t\!\in\!\mathbb{R}^{T}$, $\varpi_j\!\in\!\mathbb{R}^{T}$ and $W^j\!\in\!\mathbb{R}^{T\!\times\! T}$ are given as
\begin{equation}
\varphi^{ij}_t\!\mathrel{\mathop:}=\!\big[\langle s_i^\intercal\varPhi^{t-1}D\rangle_j,\!\langle s_i^\intercal\varPhi^{t-2}D\rangle_j,\!\cdots\!,\!\langle s_i^\intercal D\rangle_j,\!0,\!\cdots\!\big],
\end{equation}
\begin{equation}
\begin{aligned}
\varpi_j\mathrel{\mathop:}=\bigg[&\Big\langle\frac{w_\text{max}(0)-w_\text{min}(0)-2\bar{w}(0)}{w_\text{max}(0)-w_\text{min}(0)}\Big\rangle_j,\cdots,\\
&\Big\langle\frac{w_\text{max}(T\!-\!1)\!-\!w_\text{min}(T\!-\!1)\!-\!2\bar{w}(T\!-\!1)}{w_\text{max}(T\!-\!1)-w_\text{min}(T\!-\!1)}\Big\rangle_j\bigg]^\intercal, 
\end{aligned}
\end{equation}
\begin{equation}
W^j\!\!\mathrel{\mathop:}=\!\!\begin{bmatrix}
\langle \frac{w_\text{max}(0)\!-\!w_\text{min}(0)}{2}\rangle_j & & \\
& \!\ddots\! & \\
& & \langle \frac{w_\text{max}(T\!-\!1)\!-\!w_\text{min}(T\!-\!1)}{2}\rangle_j
\end{bmatrix},
\end{equation}
where $\langle a\rangle_i$ denote the $i$-th component in vector $a$. The support function in (\ref{equ:statesup}) is the dual of infinity norm, expressed as
\begin{equation}
\|(\varphi^{ij}_t W^j)^\intercal\|_\infty^\ast\mathrel{\mathop:}=\sup_{\tilde{w}}\{\varphi^{ij}_t W^j \tilde{w}:\|\tilde{w}\|_\infty\leq 1\},
\end{equation}
where $\|\cdot\|^\ast_\infty$ denotes the dual of infinity norm. It has been proven by H\"{o}lder's inequality \cite{boyd2004convex} that the dual of $p$-norm is $q$-norm where $\frac{1}{p}\!+\!\frac{1}{q}\!=\!1$. Thus, $\|(\varphi^{ij}_t W^j)^\intercal\|_\infty^\ast=\|(\varphi^{ij}_t W^j)^\intercal\|_1$.
Finally, tightened constraints $\bar{\mathbb{X}}(t),\!t\!\in\!\!\mathbb{I}_1^T$ can be determined by
\begin{equation}
\label{equ:tightx}
\begin{aligned}
\bar{\mathbb{X}}(t)\!\mathrel{\mathop:}=\!\Big\{\!x\!\in\!\mathbb{R}^{n_x}\!\!: &s^\intercal_i x\leq r_i-\\
&\sum_{j=1}^{n_w}\! (\!\|\varphi^{ij}_t W^j \|_1\!\!-\!\varphi^{ij}_t W^j\varpi_j\!),\!\forall i\!\in\!\mathbb{I}_1^M\!\Big\}.
\end{aligned}
\end{equation}
Note that $\sum_{j=1}^{n_w} (\|\varphi^{ij}_t W^j \|_1\!-\!\varphi^{ij}_t W^j\varpi_j)$ indicates the ``degree'' of bound restriction for the $i$-th state constraint, and is calculated by simple algebraic operations without iteratively solving linear programming.

\subsection{Conservativeness Reduction Using Budget Uncertainty Set}
The box uncertainty set models disturbances in a conservative way, since uncertainties unlikely turn out to be their worst values simultaneously. Here the budget uncertainty set \cite{bertsimas2004price} is combined with multi-period constraint tightening to flexibly adjust the conservativeness level. The budget uncertainty set induces tight bounds on sums of random variables, given by
\begin{equation}
\mathcal{W}=\mathcal{W}_1\cap\mathcal{W}_\infty\mathrel{\mathop:}=\left\{\tilde{w}:\|\tilde{w}\|_1\leq\Gamma,\|\tilde{w}\|_\infty\leq 1\right\},
\end{equation}
where $\Gamma$ is called the ``budget'' of uncertainty. The tightened constraint $\bar{\mathbb{X}}(t)$ for the budget uncertainty set is then given by
\begin{equation}
\label{equ:conbudget}
\begin{aligned}
\bar{\mathbb{X}}(t)\!\mathrel{\mathop:}=\!\Big\{&x\in\mathbb{R}^{n_x}:s^\intercal_i x\leq r_i-\\
&\sum_{j=1}^{n_w} \sup_{\tilde{w}}\{\varphi^{ij}_t W^j (\tilde{w}\!+\!\varpi_j)\!:\!\tilde{w}\!\in\!\mathcal{W}\},\!\forall i\!\in\!\mathbb{I}_1^M\Big\}.
\end{aligned}
\end{equation}
A tractable formulation of (\ref{equ:conbudget}) can be derived based on the following property of support functions \cite{borwein2010convex}.

\emph{Lemma 1}. Let $\mathcal{W}_1,\!\cdots\!,\mathcal{W}_k$ be closed convex sets, and $\mathcal{W}\!=\!\cap_{i=1}^k\mathcal{W}_i$. If $\cap_{i=1}^k\text{relint}(\mathcal{W}_i)\neq\emptyset$, then for a given vector $y$,
\begin{equation}
\sup_{w\in\mathcal{W}}y^\intercal w=\min_{y_1,\cdots,y_k}\Big\{\sum_{i=1}^k\sup_{w\in\mathcal{W}_i}y_i^\intercal w:\sum_{i=1}^k y_i=y\Big\}.
\end{equation}

According to Lemma 1,
\begin{equation}
\begin{aligned}
&\sup_{\tilde{w}}\big\{\varphi^{ij} W^j \tilde{w}:\tilde{w}\in\mathcal{W}\big\} \\
=&\min_{y_1,y_\infty}\!\big\{\!\!\sup_{\tilde{w}\in\mathcal{W}_1}\!\! y_1^\intercal\tilde{w}\!+\!\!\!\sup_{\tilde{w}\in\mathcal{W}_\infty}\!\!y_\infty^\intercal\tilde{w}:y_1\!+\!y_\infty\!=\!(\varphi^{ij} W^j)^\intercal\big\} \\
=&\min_{y_1,y_\infty}\big\{\Gamma\|y_1\|_1^\ast+\|y_\infty\|_\infty^\ast:y_1+y_\infty=(\varphi^{ij} W^j)^\intercal\big\} \\
=&\min_{y_1,y_\infty}\big\{\Gamma\|y_1\|_\infty+\|y_\infty\|_1:y_1+y_\infty=(\varphi^{ij} W^j)^\intercal\big\}\\
=&\min_{y_1}\big\{\Gamma\|y_1\|_\infty+\|(\varphi^{ij} W^j)^\intercal-y_1\|_1\big\}.
\end{aligned}
\end{equation}
Then $\bar{\mathbb{X}}(t)$ for the budget uncertainty can be determined by,
\begin{equation}
\label{equ:onestep}
\begin{aligned}
\bar{\mathbb{X}}(t)\!\mathrel{\mathop:}=\!\Big\{x\!\in\!\mathbb{R}^{n_x}: &s^\intercal_i x\leq r_i-\\
&\sum_{j=1}^{n_w}(\gamma^{ij}(\Gamma)\!-\!\varphi^{ij}_t W^j\varpi_j),\forall i\!\in\!\mathbb{I}_1^M\Big\},
\end{aligned}
\end{equation}
where $\gamma^{ij}(\Gamma) = \min_{y_1}\big\{\Gamma\|y_1\|_\infty + \|(\varphi^{ij} W^j)^\intercal -y_1\|_1\big\}$.
Equation (\ref{equ:onestep}) explicitly characterizes the impact of uncertainties on constraint restrictions.

\vspace{-4mm}
\section{Simulation Results}
\vspace{-2mm}
\subsection{System Configuration}
\begin{figure}[htb]
\centering
\includegraphics[width=0.99\columnwidth]{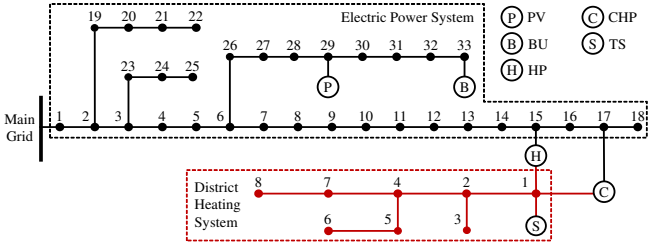}
\caption{The system diagram of CHPS}
\label{fig:CHPS}
\end{figure}

\begin{figure}[htb]
	\centering
	\subfloat[Net electricity loads.]{\includegraphics[width=0.49\columnwidth]{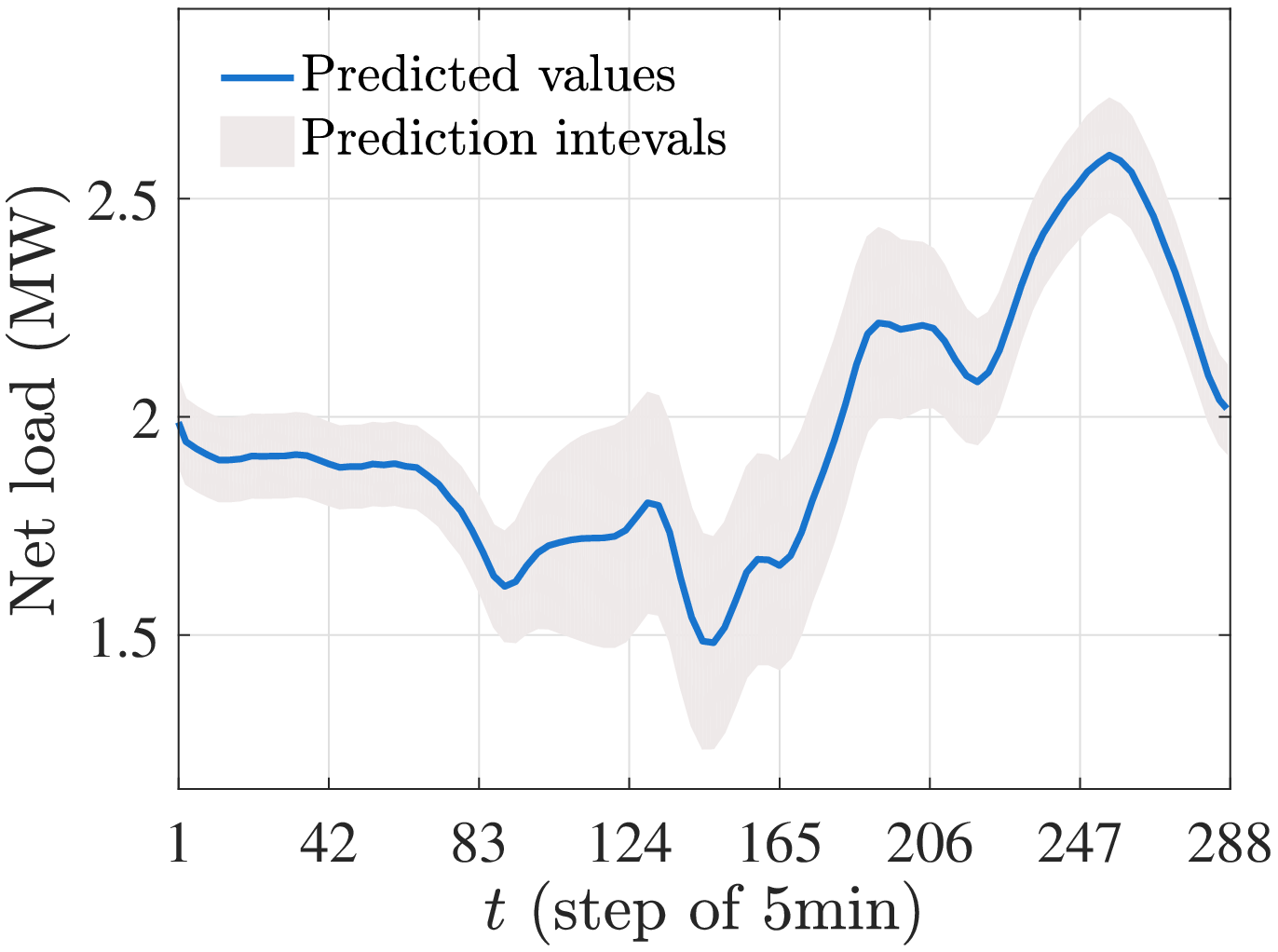}
		\label{fig:NetLoad}}
	\subfloat[Total heating loads.]{\includegraphics[width=0.49\columnwidth]{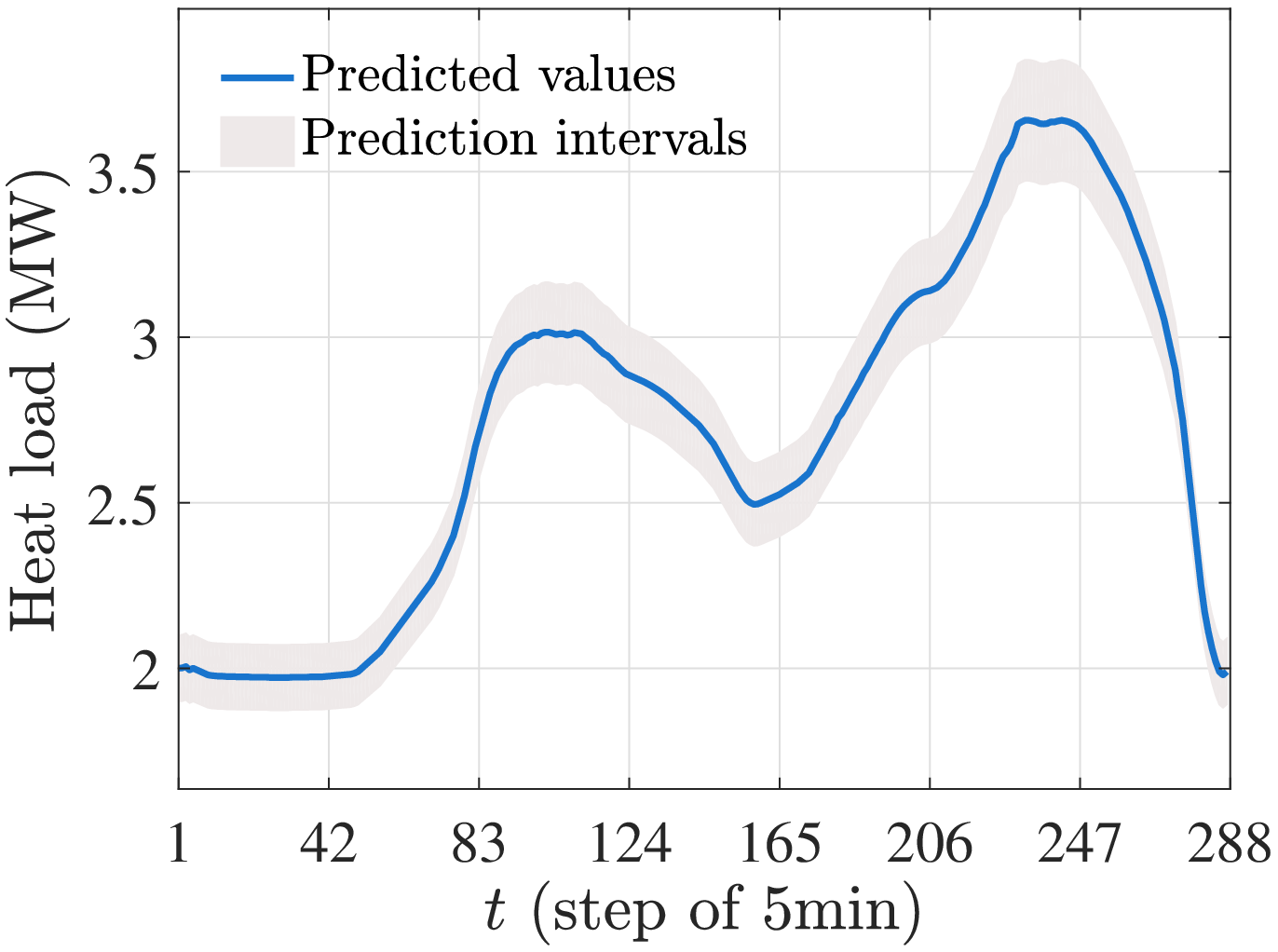}
		\label{fig:HeatLoad}}
	\caption{Profiles of net electricity loads and thermal loads.}
	\label{fig:Load}
\end{figure}

\vspace{-2mm}
The test system based on a 33-bus distribution network \cite{baran1989network} and an 8-node district heating network \cite{li2016combined}, as shown in Fig \ref{fig:CHPS}, is utilized to verify the proposed ERD approach. 
Typical profiles of net electricity and heat loads in a winter day are shown in Fig. \ref{fig:Load}. Simulation results are obtained based on a PC with Intel Core i7-8750H @2.2GHz, 16GB RAM. Algorithms are tested using MATLAB R2016b with YALMIP \cite{Lofberg2004}. 

\vspace{-4mm}
\subsection{Effectiveness of ERD for CHPS}
\begin{figure}[htb]
	\centering
	\subfloat[Trajectory of state of charge.]{\includegraphics[width=0.49\columnwidth]{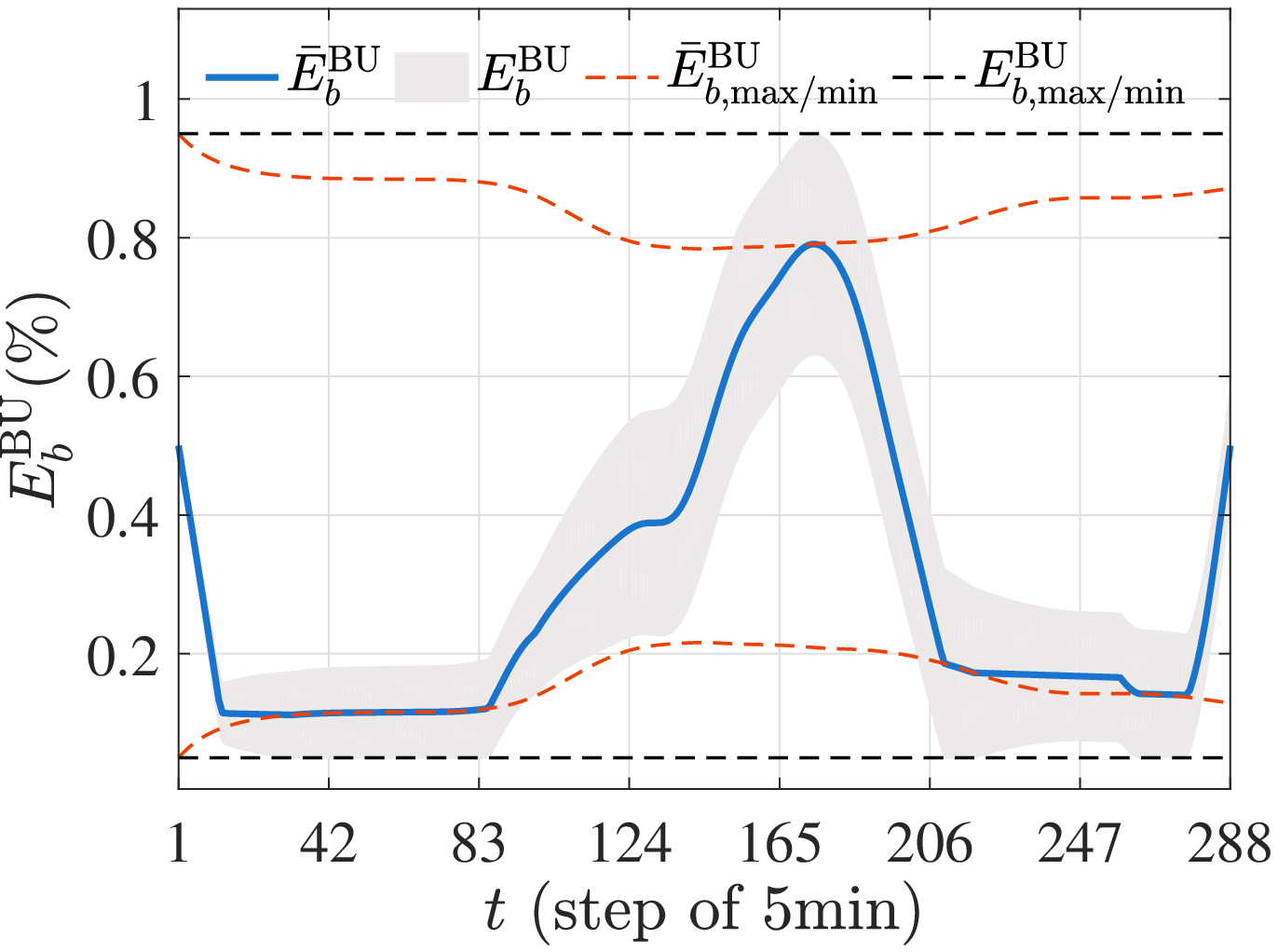}
		\label{fig:RCLCBoxSOC}} 
	\subfloat[Trajectory of heat storage level.]{\includegraphics[width=0.49\columnwidth]{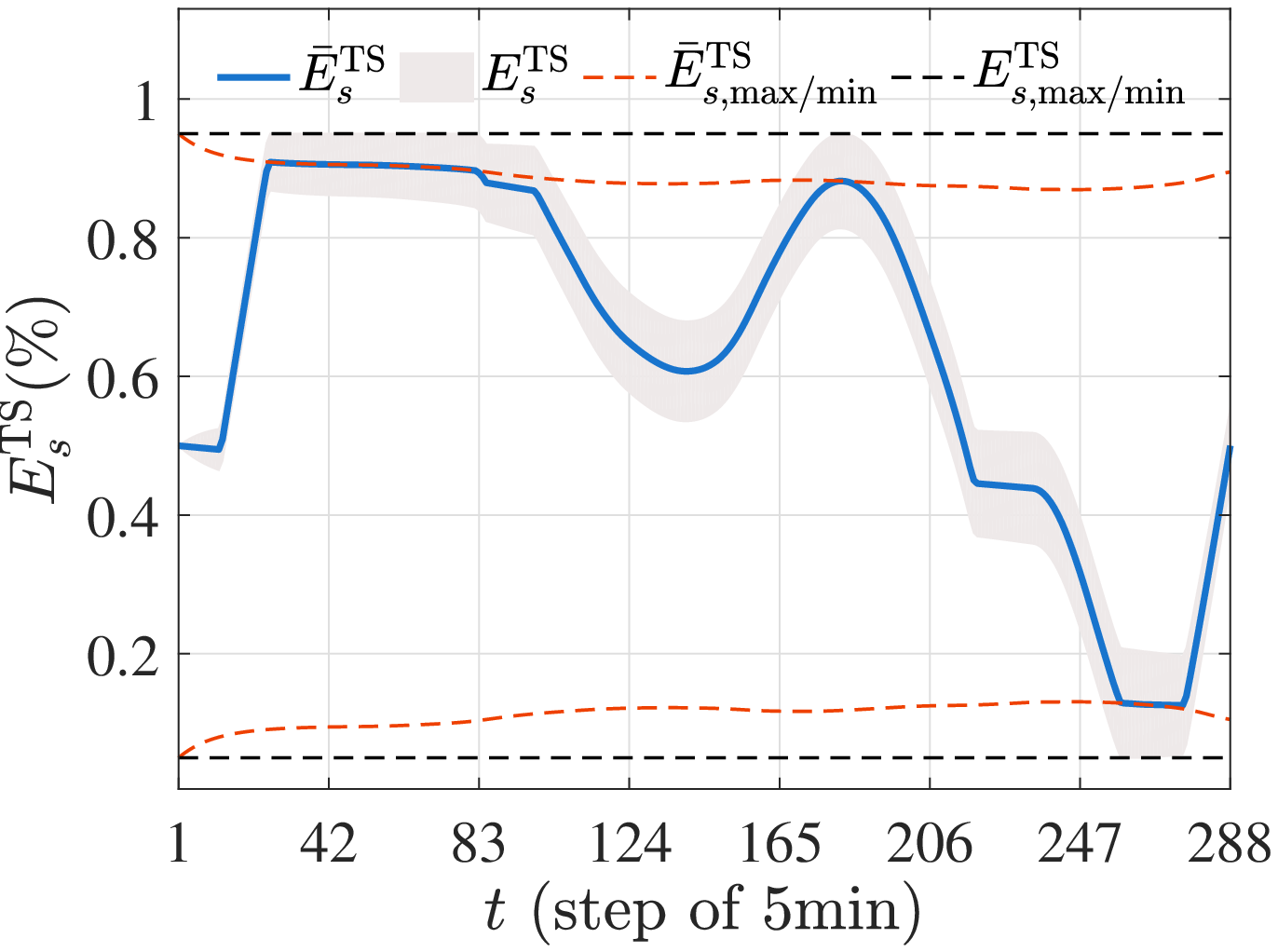}
		\label{fig:RCLCBoxTSL}} \\
	\subfloat[Power output of the CHP.]{\includegraphics[width=0.49\columnwidth]{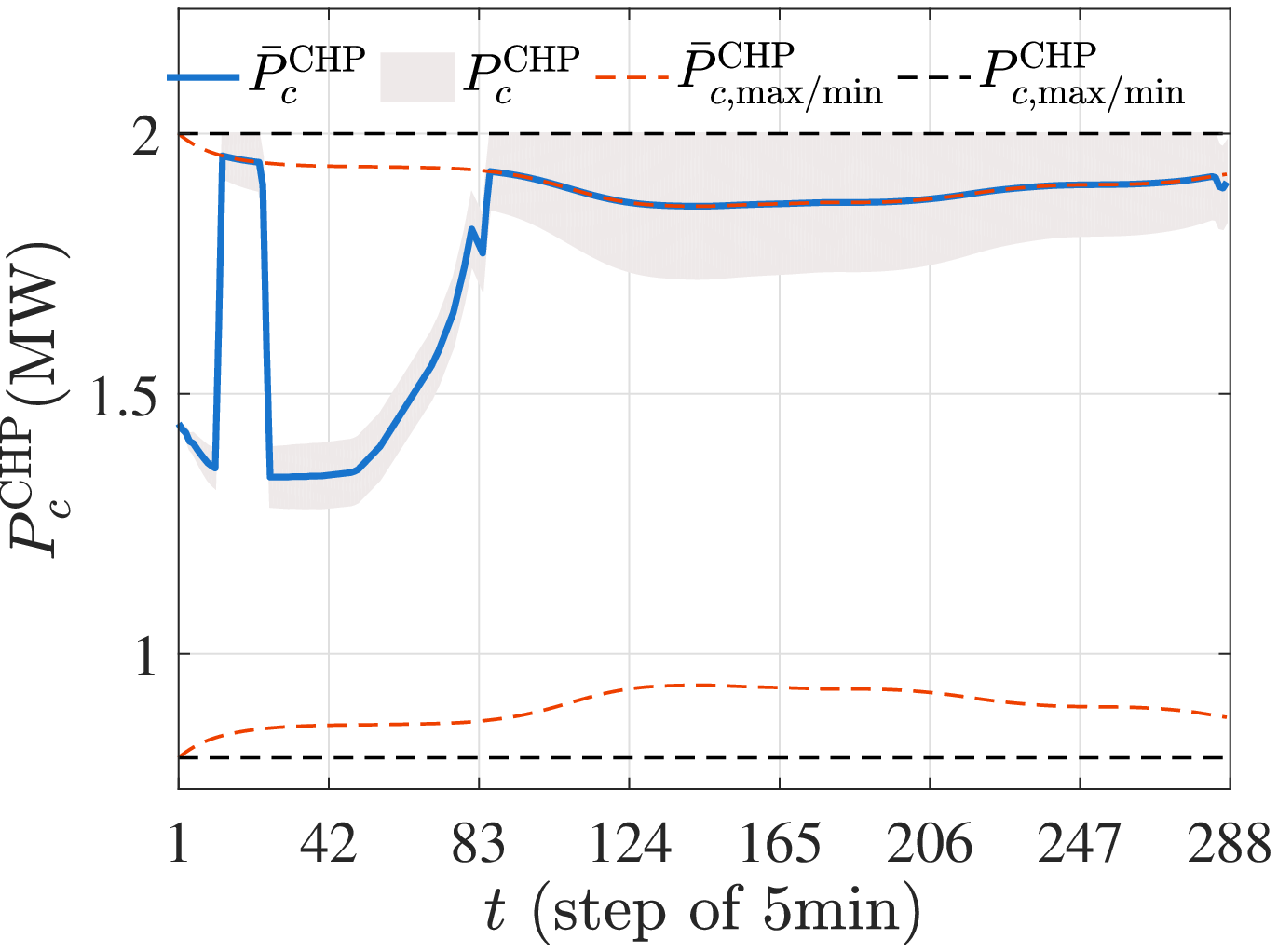}
		\label{fig:RCLCBoxCHP}}
	\subfloat[Exchange power with main grid.]{\includegraphics[width=0.49\columnwidth]{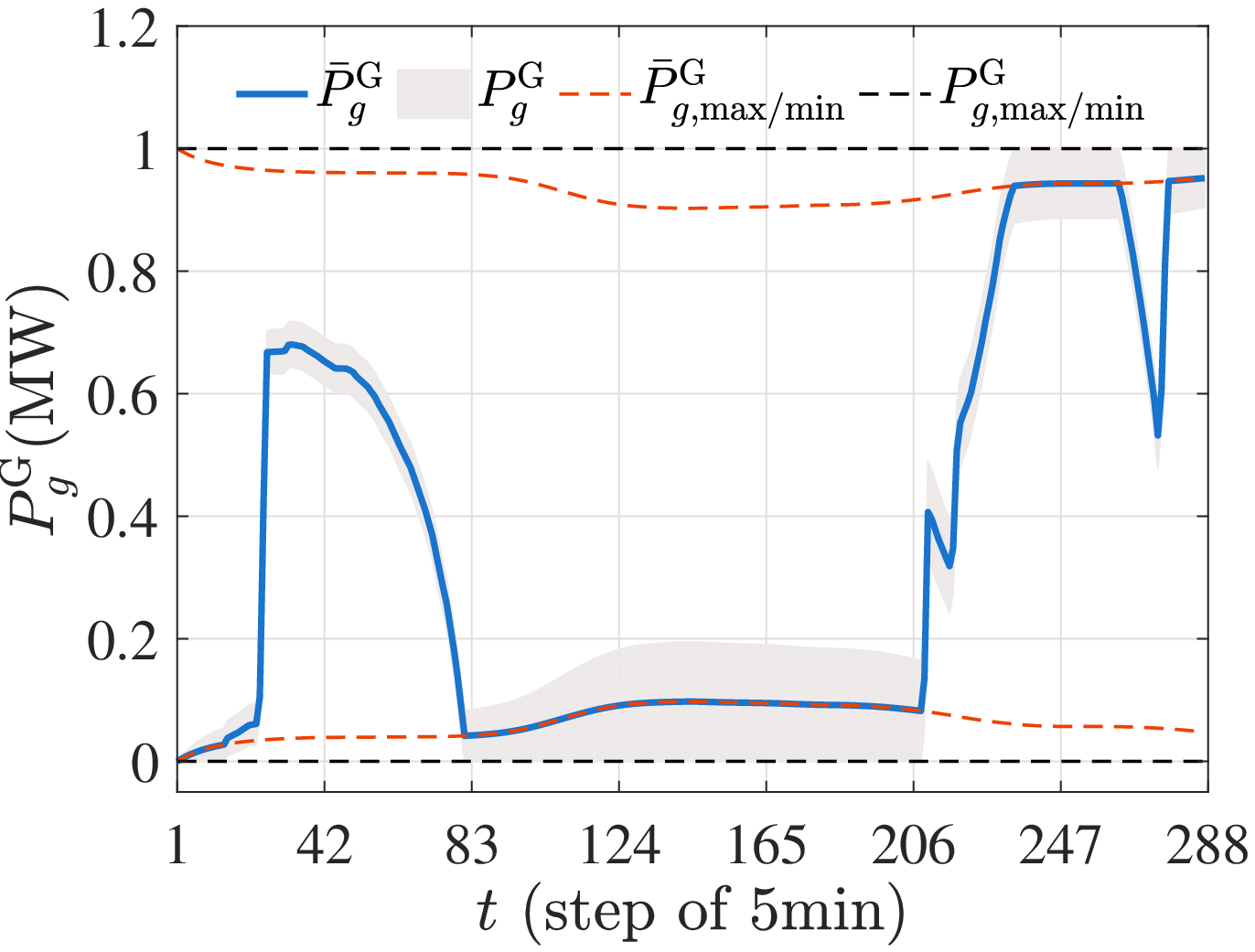}
		\label{fig:RCLCBoxGrid}}
	\caption{Dispatch strategy from the ERD model.}
	\label{fig:RCLCBox}
\end{figure}

\begin{figure}[htb]
	\centering
	\subfloat[Trajectory of heat storage level.]{\includegraphics[width=0.49\columnwidth]{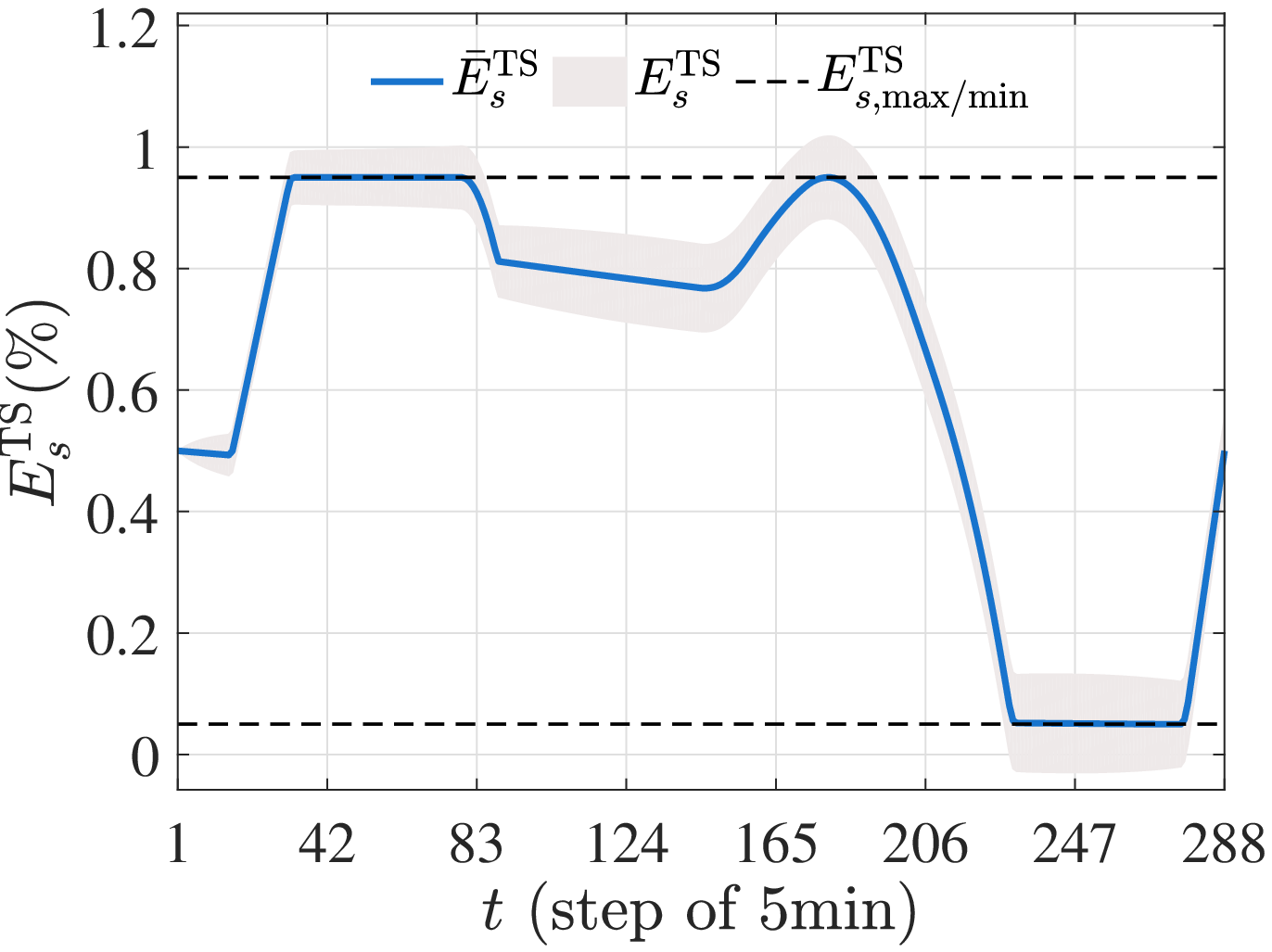}
		\label{fig:NomTSL}}
	\subfloat[Power output of the CHP.]{\includegraphics[width=0.49\columnwidth]{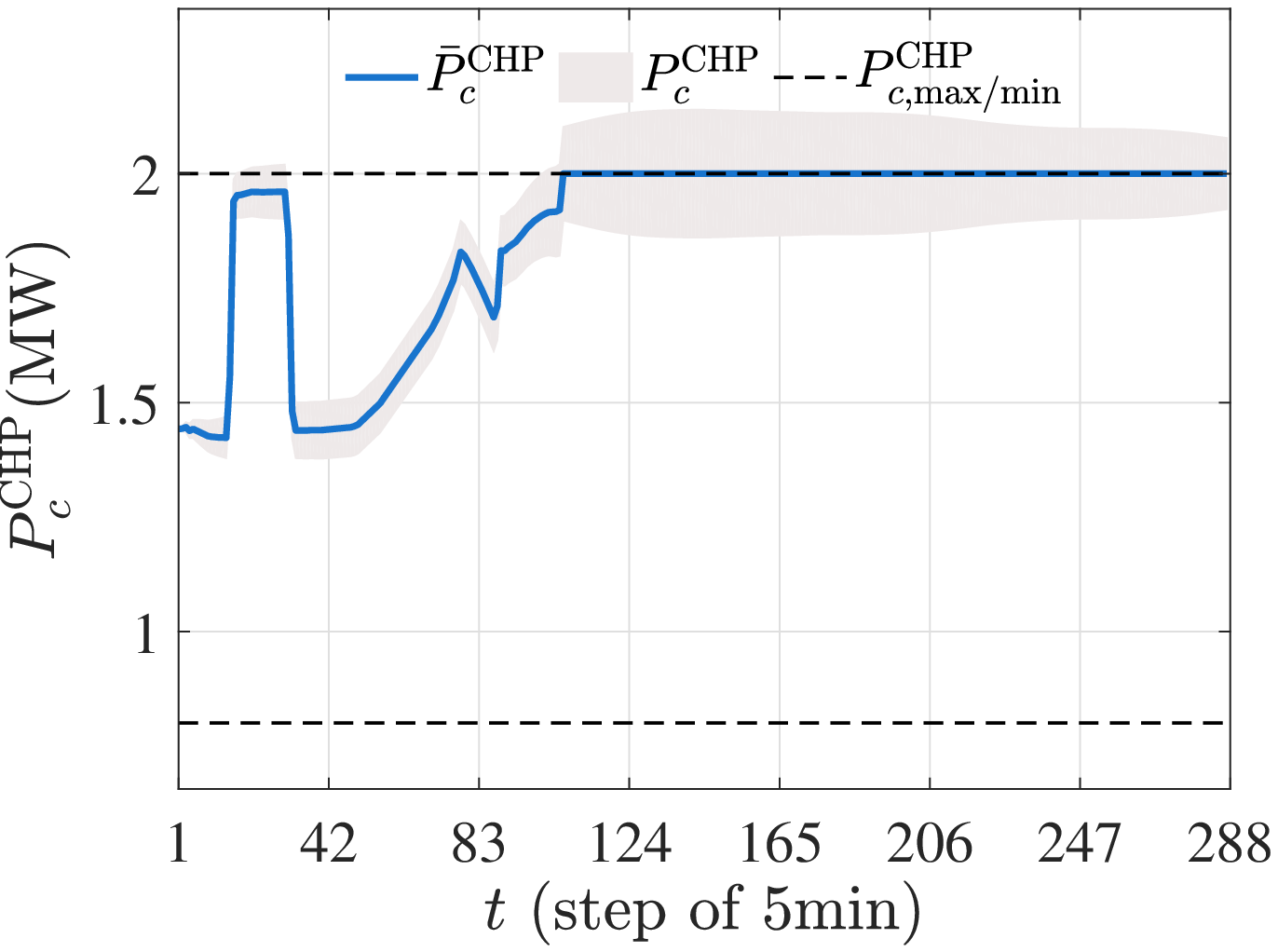}
		\label{fig:NomCHP}}
	\caption{Dispatch strategy from the deterministic model.}
	\label{fig:Nom}
\end{figure}

\vspace{-2mm}
A 24-hour operational strategy (with 5-minute dispatch intervals) obtained from the ERD method is demonstrated in Fig. \ref{fig:RCLCBox}, where
the robustness of the proposed ERD approach is explicitly verified. Solid blue lines indicate nominal values of control inputs and states, and possible deviations of state and control variables induced by disturbance effects are covered by shaded areas. 
Due to the multi-period tightened constraints (shown as red dashed lines), the robustness is guaranteed effectively since all possible realizations of state and control variables are enveloped within the original constraints (denoted as black dashed lines). For instance, the nominal electric output of CHP at $t\!=\!120$ is restricted in $0.928\!\leq\!\bar{P}^\text{CHP}_c\!\leq\!1.872$ to avoid violation of the original constraint $0.8\!\leq\! P^\text{CHP}_c\!\leq\! 2.0$. On the contrary, solving a deterministic dispatch problem without tightened constraints leads to constraint violations, as illustrated in Fig. \ref{fig:Nom}, and hence cannot guarantee the robustness of operational strategies. In Fig. \subref*{fig:NomTSL}, the thermal energy level may violate the restriction of $0\%\!\leq\!E^\text{TS}_s\!\leq\!100\%$. The CHP also operates at overloading conditions in Fig. \subref*{fig:NomCHP}. In addition, the degree of constraint restriction at noontime is higher than other periods, since increasing PV power brings more uncertainties especially in the electric power sector.

Computing time for determining multi-period tightened constraints is given in Table \ref{tab:tightened} using the proposed direct constraint tightening algorithm and the iteration method \cite{kolmanovsky1998theory}. The dispatch horizon is one day with a time step of 5 minutes. Hence, 288-fold tightened constraints must be derived. Generating tightened constraints using the proposed algorithm with box uncertainty only requires simple algebraic calculations. In contrast, the iteration method needs to repeatedly solve linear programming problems to determine constraint restrictions at each time instant. Thus, the computation time of the proposed method drops significantly from hours to less than 1 second. For the budget uncertainty set, the proposed method is faster by two orders of magnitude due to less time of solving linear programming problems compared to the iterative solution.
\begin{table}[htb]
	\footnotesize
	\centering
	\caption{Computing Time of Determining Tightened Constraints}
	\label{tab:tightened}
	\begin{tabular}{@{\hspace{4pt}}c@{\hspace{6pt}}c@{\hspace{6pt}}c@{\hspace{6pt}}c@{\hspace{6pt}}c@{\hspace{6pt}}c@{\hspace{6pt}}c@{\hspace{4pt}}}
		\toprule 
		\multirow{2}{*}{Method} & \multirow{2}{*}{Uncer. type} & \multicolumn{5}{c}{Computing time for each type of constraints (s)} \\
								& 		  & $\bar{\mathbb{X}}(t)$ & $\bar{\mathbb{U}}(t)$ & $\bar{\mathbb{Y}}(t)$ & $\Delta \bar{\mathbb{U}}(t)$ & $\Delta \bar{\mathbb{Y}}(t)$ \\ \midrule
		\multirow{2}{*}{Iteration} & Box & 7048.13 & 11977.35 & 80446.64 & 11527.43 & 8220.56 \\
								   & Budget & 9569.45 & 14338.48 & 95632.16 & 12786.70 & 10892.59 \\ \midrule
		\multirow{2}{*}{Direct} & Box & 0.0121 & 0.0160 & 0.2278 & 0.0157 & 0.0156 \\
								& Budget & 92.44 & 177.66 & 1417.30 & 156.90 & 104.75 \\ \bottomrule				   
	\end{tabular}
\end{table}

\vspace{-4mm}
\subsection{Validation of Budget Uncertainty Set}
\begin{figure}[htb]
	\centering
	\subfloat[Trajectory of state of charge.]{\includegraphics[width=0.49\columnwidth]{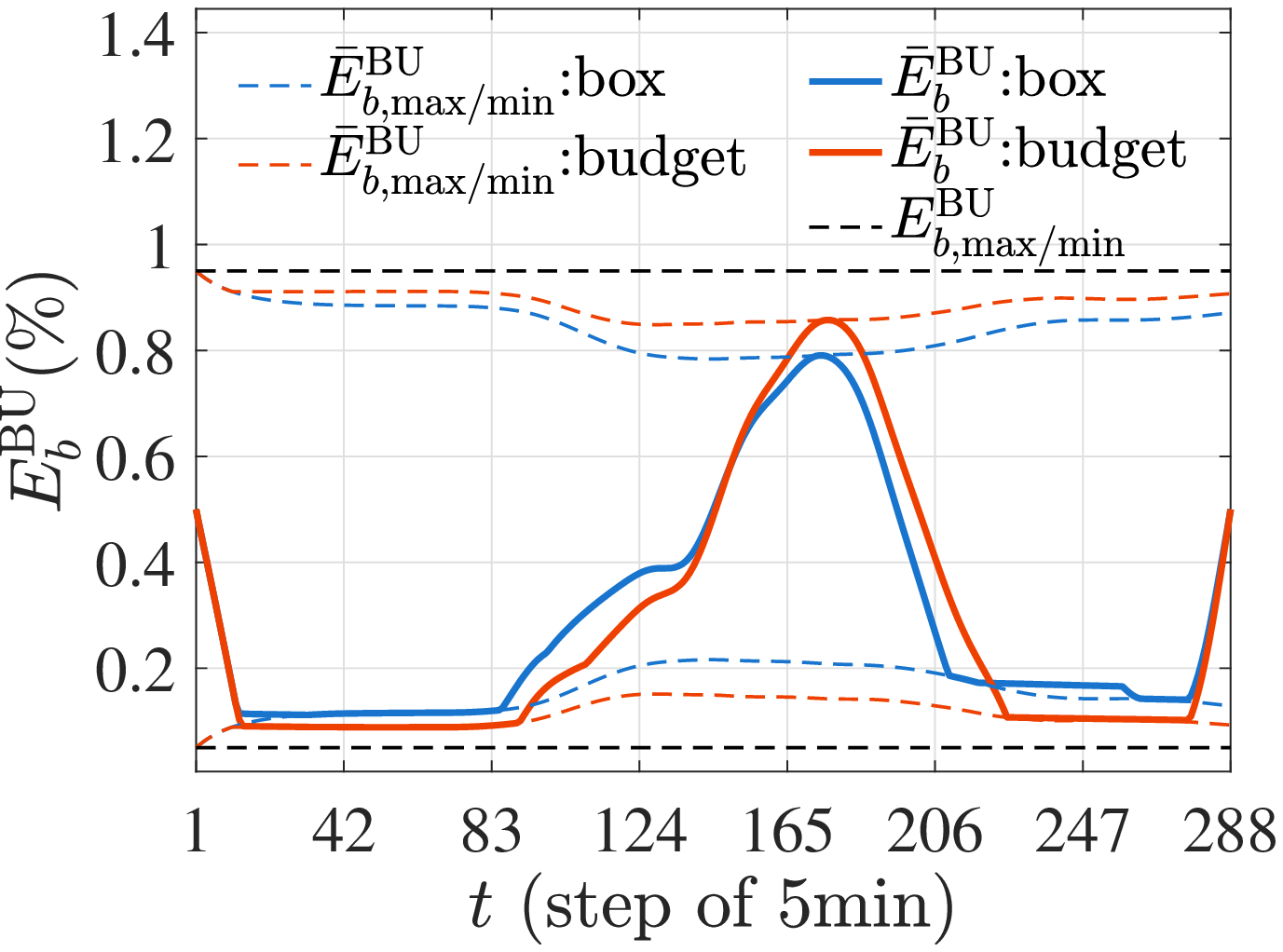}
		\label{fig:SOCCompare}}
	\subfloat[Trajectory of heat storage level.]{\includegraphics[width=0.49\columnwidth]{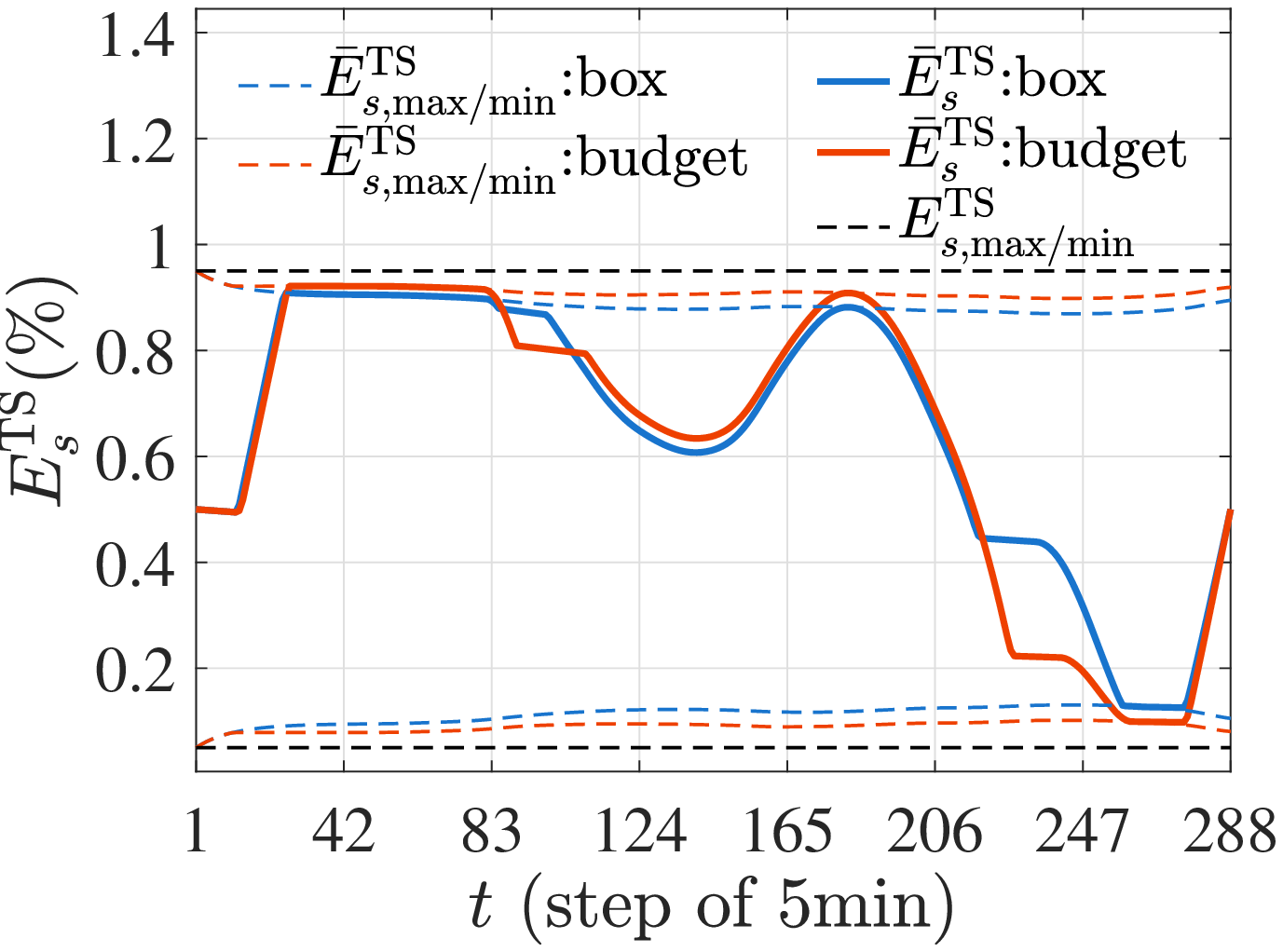}
		\label{fig:TSLCompare}}
	\caption{Dispatch results of ERD model with box and budget uncertainties.}
	\label{fig:BoxBudgetCompare}
\end{figure}

\vspace{-2mm}
Fig. \ref{fig:BoxBudgetCompare} shows nominal dispatch results from the ERD model with box and budget uncertainty sets ($\Gamma\!=\!10$). By incorporating the budget uncertainty set, state constraint restrictions, indicated as red dashed lines in Fig. \subref*{fig:SOCCompare} and Fig. \subref*{fig:TSLCompare}, are less conservative compared to those of box uncertainties. As a consequence, the evolution of the nominal energy storage levels spans a larger area under the budget uncertainty. 

\begin{figure}[htb]
\centering
\includegraphics[width=0.50\columnwidth]{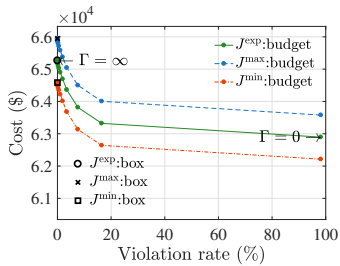}
\caption{Multiple solutions generated with different budgets.}
\label{fig:TradeOff}
\end{figure}

The price of robustness can be controlled by budget $\Gamma$ to achieve better trade-offs between constraint satisfaction and cost reduction. Increasing $\Gamma$ generally enlarges the uncertainty set, and leads to a more conservative solution with higher operational costs and a lower probability of constraint violation. 
To examine the quality of the robust dispatch strategy, 1,000,000 samples of the uncertainty variables are generated randomly based on the predicted intervals to derive several metrics, including the constraint violation rate, expected operational cost, and maximum and minimum operational costs. Fig. \ref{fig:TradeOff} gives the solutions of different budgets with respect to the constraint violation rate and operational cost, as well as the ERD solution with box uncertainty. The expected cost drops from \$65265 to \$64362 with a budget of $\Gamma\!=\!10$ and 3.5\% of constraint violation. If $\Gamma\!=\!2$, a total reduction of \$1943 is achieved with a violation rate of 16.5\%. In general, costs are reduced by sacrificing a relatively small level of robustness.

\vspace{-4mm}
\subsection{Performance Comparison with Min-Max Algorithm}
\vspace{-2mm}
This section presents comparisons among the proposed ERD method, min-max robust optimization (RO) and deterministic optimization (DO). Specifically, RO is applied with different linear decision rules, including fixed decision rules (RO-Fix) and variable decision rules (RO-Var). RO-Fix utilizes constant linear coefficients to map control inputs at $t$ with disturbance at $t\!-\!1$, while RO-Var uses time-variant coefficients.

The computational complexity is presented in Table \ref{tab:complexity} by comparing the proposed ERD, RO and DO methods. The ERD formulation excludes uncertainties without introducing auxiliary variables. Thus, the computation complexity of ERD is identical to the DO model. The average computing time of ERD grows slightly from 1.0479s to 8.2871s when the dispatch interval is switched from one hour to 5 minutes. On the contrary, the RO needs to introduce robust counterparts with additional decision variables. The problem size and computing time grow tremendously when going from hourly to 5-min dispatch intervals, which is computationally insufficient for multi-period dispatch. For instance, the number of variables for RO-Fix is approximately 1000 times higher compared with the ERD model in the 288-time period dispatch problem. Average computation time for RO-Fix rises from 1.4901s to 778.96s due to a remarkable increase of the problem size. Furthermore, the utilization of variable decision rules intensifies the computational issue with a further increase in the number of variables. The proposed ERD method is proved to have better scalability for multi-period dispatch problems.
\begin{table}[htb]
	\footnotesize
	\centering
	\caption{Computational Complexity of Different Methods}
	\label{tab:complexity}
	\begin{tabular}{@{\hspace{7pt}}c@{\hspace{7pt}}c@{\hspace{7pt}}c@{\hspace{7pt}}c@{\hspace{7pt}}c@{\hspace{7pt}}}
		\toprule 
		Item & RO-Fix & RO-Var & DO & ERD \\ \midrule
		\multicolumn{5}{c}{$T=24,\Delta T=60$min} \\ \midrule
		\# of variables & 7398 & 7524 & 869 & 869 \\
	    \# of equal. constr. & 1633 & 1633 & 687 & 687 \\ 
	    \# of inequal. constr. & 12794 & 12794 & 801 & 801 \\ 
	    Aver. CPU time (s) & 1.4901 & 1.8699 & 1.0060 & 1.0479 \\ 	\midrule
	    \multicolumn{5}{c}{$T=96,\Delta T=15$min} \\ \midrule
	    \# of variables & 72126 & 72684 & 3605 & 3605 \\
	    \# of equal. constr. & 20425 & 20425 & 2847 & 2847 \\ 
	    \# of inequal. constr. & 108770 & 108770 & 3321 & 3321 \\ 
	    Aver. CPU time (s) & 25.1901 & 30.2094 & 2.5612 & 2.6083 \\ 	\midrule  
	    \multicolumn{5}{c}{$T=288,\Delta T=5$min} \\ \midrule
	    \# of variables & 11713222 & 11714932 & 10901 & 10901 \\
	    \# of equal. constr. & 171913 & 171913 & 8607 & 8607 \\ 
	    \# of inequal. constr. & 770210 & 770210 & 10041 & 10041 \\ 
	    Aver. CPU time (s) & 778.96 & 950.01 & 8.0372 & 8.2871 \\ 
	    \bottomrule
	\end{tabular}
\end{table}

The dispatch performance of different algorithms is evaluated using 1,000,000 random simulations based on multiple criteria, including the nominal cost, expected cost, maximum and minimum cost, and the probability of constraint violation. Detailed numerical results are listed in Table \ref{tab:performance}. The DO method produces an economically efficient strategy, but robustness is quite low with nearly 100\% constraint violation. The solution of ERD with box uncertainty is robust, and performs better than RO-Fix with reduction of the expected operational cost from \$65796 to \$65265. If variable linear decision rules are utilized, RO-Var performs better than ERD-box with the expected cost of \$64927. However, extensive computational effort is required by RO-Var (Table \ref{tab:complexity}). In contrast, ERD with budget uncertainties ($\Gamma\!=\!10$) produces a more economical solution (\$64362 in expectation) with flexibility in terms of conservativeness. Overall, the ERD method achieves good dispatch performance with substantially higher computational efficiency and adjustable conservativeness level compared with the min-max RO strategy.
\begin{table}[htb]
	\footnotesize
	\centering
	\caption{Comparison on Dispatch Performance}
	\label{tab:performance}
	\begin{tabular}{@{\hspace{1pt}}c@{\hspace{5pt}}c@{\hspace{5pt}}c@{\hspace{5pt}}c@{\hspace{5pt}}c@{\hspace{5pt}}c@{\hspace{5pt}}c@{\hspace{1pt}}}
		\toprule 
		Item  & DO & RO-Fix & RO-Var & ERD-box & ERD-$\Gamma\!=\!10$ \\ \midrule
		$J^\text{nom}$ (\$) & 62876 & 65765 & 64893 & 65247 & 64346 \\ 
		$J^\text{exp}$ (\$) & 62895 & 65796 & 64927 & 65265 & 64362 \\ 
		$J^\text{max}$ (\$) & 63578 & 66304 & 65511 & 65947 & 65044 \\ 
		$J^\text{min}$ (\$) & 62212 & 65325 & 64428 & 64584 & 63679 \\ 
		Prob. (\%) & 98.18 & 0 & 0 & 0 & 3.47 \\ 
		\bottomrule
	\end{tabular}
\end{table}

\vspace{-4mm}
\section{Conclusion}
\vspace{-2mm}
This paper proposes a novel efficient robust dispatch model of combined heat and power systems based on extensions of disturbance invariant sets, which has high computational efficiency and enables flexible adjustments in the conservativeness level of the resulting operational strategies. The proposed ERD model achieves robustness against uncertainties by solving a nominal uncertainty-free problem with multi-period tightened constraints, and preserves computational scalability in fine-grained multi-period dispatch problems. A direct constraint tightening algorithm is developed based on the dual norm to calculate constraint restrictions efficiently without iterations considering time-variant uncertainty sets. The budget uncertainty set is newly combined with constraint tightening in the proposed ERD model to reduce conservativeness levels. Besides, network-constrained electric power flow and temperature dynamics are modeled to support improved operational decisions under realistic system conditions.
Comprehensive case studies verify the dispatch robustness and computational efficiency of the proposed ERD method compared to traditional min-max robust optimization.
In summary, the ERD method facilitates robust dispatch strategies for CHPS with improved computational and economic performance.
\vspace{-4mm}


\ifCLASSOPTIONcaptionsoff
  \newpage
\fi



\bibliographystyle{IEEEtran}
\bibliography{robustMPCforIES}

\end{document}